\documentclass[11pt]{article}
\pdfoutput=1 
\usepackage{jcappub, url, enumerate, cancel}

\usepackage{amstext}
\usepackage{amsmath}
\usepackage{latexsym}
\usepackage{amsfonts}
\usepackage{amssymb}
\usepackage{color}
\usepackage{mathtools}
\usepackage{graphics}
\usepackage{graphicx}

\usepackage[normalem]{ulem}

\title{Scalar Representations in the Light of Electroweak Phase Transition
  and Cold Dark Matter Phenomenology}

\author[a]{Shehu S. AbdusSalam}
\affiliation[a]{The Abdus Salam International Center for Theoretical Physics
Strada Costiera 11, 34151, Trieste, Italy}

\author[b]{and Talal Ahmed Chowdhury}
\affiliation[b]{SISSA, Via Bonomea 265, 34136, Trieste, Italy}

\emailAdd{shehu@ictp.it}
\emailAdd{chowdhu@sissa.it}

\abstract{
The extension of the standard model's minimal Higgs sector with an 
inert $SU(2)_L$ scalar doublet can provide light dark matter candidate
and simultaneously induce a strong phase transition for explaining
Baryogenesis. There is however no symmetry reasons to prevent the
extension using scalars with higher $SU(2)_L$ representations. By
making random scans over the models' parameters, we show
that in the light of electroweak physics constraints, strong first
order electroweak phase transition  and the possibility of having
sub-TeV cold dark matter candidate the 
higher representations are rather disfavored compared to the inert
doublet. This is done by computing generic perturbativity
behavior and impact on electroweak phase transitions of higher
representations in comparison with the inert doublet model. Explicit
phase transition and cold dark matter phenomenology within the context
of the inert triplet and quartet representations are used for detailed
illustrations.}

\begin{document}

\maketitle

\section{Introduction}

The discovery of an about 126 GeV Higgs boson ~\cite{:2012gk,:2012gu}
is yet another important support for and completion of the
Standard Model (SM). The SM with minimal Higgs sector 
contains one complex Higgs doublet 
which after electroweak symmetry breaking gives a neutral CP-even 
Higgs boson. But one can consider a scenario with singlets, with more than one
doublet or with an $N$ copies of $SU(2)_L$ n-tuplets where the (non standard) Higgs sector can be
used to explicitly account for Baryogenesis and cold dark matter. Using these phenomena and related experimental data we
set to qualitatively explore for the preferred scalar representation
in the SM. 

Extensive astrophysical and cosmological observations have already put
dark matter (DM) as a constituent of the universe beyond any
doubt. Although we still have to determine which is the candidate
for the dark matter from particle physics point of view, the most
popular one is considered to be the stable weakly interacting particle
(WIMP) \cite{Scherrer:1985zt} for which the observed DM relic density
is obtained if it's mass lies near the electroweak scale. Apart from
DM identification, one other unresolved
question within SM is the observed matter-antimatter asymmetry of the
universe. Such asymmetry is described by Baryogenesis scenario first
put forward by Sakharov \cite{Sak} and one essential ingredient of
this mechanism is 'out of equilibrium process'. 

Now that the study of beyond standard model (BSM) physics is being
explored with the LHC, a well motivated scenario within the testable
reach of LHC is Electroweak Baryogenesis \cite{Kuzmin:1985mn} where out of equilibrium condition is given by
strong first order phase transition. The SM has all the tools required
by the Sakharov's condition for Baryogenesis, i.e. baryon number
violation at high temperature through sphalerons
\cite{Klinkhamer:1984di, Arnold:1987mh, Dine:1991ck}, C
and CP violation with CKM phase and strong first order phase
transition \cite{Anderson:1991zb, Dine:1992wr}. However, it was shown that to avoid baryon washout
by sphalerons, Higgs mass has to be below $45$ GeV for
strong electroweak phase 
transition (EWPhT) \cite{Shaposhnikov:1987tw, Bochkarev:1987wf, Shaposhnikov:1987pf, 
Bochkarev:1990fx}, which was later confirmed
by lattice studies \cite{Kajantie:1995kf, Kajantie:1996mn,
  Kajantie:1996qd} and eventually it was ruled
out by the LEP data \cite{lephiggs}. Now with Higgs at $126$ GeV, clearly one 
can see the requirement of extending SM by new particles,
possibly lying nearly the electroweak scale, which could not only provide
strong EWPhT for explaining matter
asymmetry but also the DM content of the universe.

One promising way is to extend the scalar sector of the SM. Within the
literature there are numerous considerations for non minimal Higgs
sector with various representations of the additional Higgs multiplet 
in order to account for
Baryogenesis and/or dark matter. For instance, in the inert doublet
case considered in \cite{Chowdhury:2011ga, Borah:2012pu, Gil:2012ya},
it was shown that it can enable one to achieve strong EWPhT with DM 
mass lying between $45$ GeV and $80$ GeV
and predicts a lower bound on the direct detection that is consistent with XENON
direct detection limit \cite{Aprile:2012nq} (in case of sub-dominant DM, 
see \cite{Cline:2013bln}). 
Inert doublet is  a well motivated and minimal extension of scalar sector which was
first proposed as dark matter \cite{Deshpande:1977rw}, was studied as
a model for radiative neutrino mass generation \cite{Ma:2006km},
improved naturalness \cite{Barbieri:2006dq} and follows naturally
\cite{Melfo:2011ie, Martinez:2011ua} in case of mirror families
\cite{GellMann:1976pg, Wilczek:1981iz, Senjanovic:1984rw,
  Bagger:1984rk} that was to fulfill Lee and Yang's dream to restore parity \cite{Lee:1956qn}. 
The DM phenomenology regarding Inert
doublet model has been studied extensively in Refs.
\cite{LopezHonorez:2006gr,Andreas:2008xy,Hambye:2007vf,Dolle:2009fn,
LopezHonorez:2010tb,Agrawal:2008xz,Andreas:2009hj,Nezri:2009jd,
Cao:2007rm,Lundstrom:2008ai,Dolle:2009ft,Gustafsson:2010zz,
Gustafsson:2012aj,Aoki:2013lhm, Arhrib:2013ela}. Moreover, 
tentative $130$ GeV gamma line from galactic center can be accommodated
in inert doublet framework \cite{Gustafsson:2007pc}. Therefore, one
can extend scalar sector by introducing higher inert representation
and explore the nature of phase transition, consistency of the theory
at high scale and dark matter phenomenology. In this paper, we have
tried to carry out such analysis. 

Apart from the doublet, one other possibility is the scalar singlet
\cite{Profumo:2007wc, Ahriche:2007jp, Espinosa:2011ax, Cline:2012hg, Cline:2013gha} (and references there) 
which can be accounted for strong EWPhT and light DM candidate but not
simultaneously (for exceptions, see \cite{Barger:2008jx,
  Gonderinger:2012rd, Ahriche:2012ei}).\footnote{Scalar singlet can also be a force
  carrier between SM and dark matter sector inducing strong
  EWPhT \cite{Das:2009ue, Carena:2011jy} or trigger EWPhT 
  independent of being DM \cite{Ahriche:2013zwa}.} In case of larger
representations, a systematic study for DM candidate has been performed
in \cite{Cirelli:2005uq} from doublet to 7-plet of $SU(2)_L$ with both fermionic and scalar DM where
only allowed interactions of DM are gauge interactions. Additionally, scalar multiplet allows 
renormalizable quartic couplings with
Higgs doublet. In \cite{Hambye:2009pw}, study of DM phenomenology for such
scalar multiplet was carried out for large odd dimensional and real
representation and the mass of the DM turned out to be larger than the scale relevant for strong EWPhT. Besides, fermions
with large yukawa couplings to Higgs can trigger strong EWPhT by
producing large entropy when they decouple and also they can be viable
dark matter candidates \cite{Carena:2004ha} but it requires some fine
tuning of Higgs potential. Another approach with vector-like fermions
is explored in \cite{Davoudiasl:2012tu}.

Here we want to compare the various models extending the Higgs sector
using different representations in order to find the favored
representation. From the scalar sector extension, we want viable DM
candidates which will trigger strong 1st order EWPhT accounting
for Baryogenesis.  

The paper is organized as follows. In Sec.(\ref{zpt}) we present the
potential, zero temperature mass spectrum, constraints on the size of
the scalar multiplets from perturbativity and electroweak precision bounds.
In Sec.(\ref{ewpht}) we present the finite temperature
effective potential and studied the nature of phase transition for
large multiplets. In Sec.(\ref{ldm}) we present the
relic density analysis and direct detection limit for light DM in
larger multiplets. Conclusions are drawn in
Sec.(\ref{con}). In the appendices we present the relevant formulas.

\section{Scalar Representations Beyond the Standard Model}\label{zpt}
\subsection{Scalar Multiplets with Cold Dark Matter Candidates\label{zpt1}}
Any scalar multiplet charged under $SU(2)_L \times U(1)_Y$ gauge group
is characterized by $(J,Y)$. The electric charge for components of the
multiplet is given by, $Q=T_3+\frac{Y}{2}$. For half integer
representation $J=n/2$, $T_3$ ranges from $-\frac{n}{2}$ to
$\frac{n}{2}$.  So the hypercharge of the multiplet needs to be,
$Y=\pm 2T_3$ for one of the components to have neutral charge and can
be considered as DM. For integer representation $n$, similar condition
holds for hypercharge.

When one considers the lightest component of the scalar multiplet as a good dark matter candidate \cite{Taoso:2007qk}, 
it's life time has to be longer than about $10^{26}$ sec which is set by current experimental 
limits on fluxes of cosmic positrons, antiprotons and $\gamma$ radiation \cite{Hambye:2010zb, Gustafsson:2013gca}.
Such limits on the lifetime of the DM requires new couplings to be extremely 
small.\footnote{For example, inert doublet can have yukawa coupling to fermions, 
$y_{S}\bar{f}f$, that can lead into it's decay. But for $m_{DM}\sim 100$ GeV 
(illustrating WIMP scale), the bound on DM lifetime, $\tau_{DM}\sim 10^{25}-10^{27}$ 
sec sets $y_{S}\sim 10^{-25}-10^{-27}$. Also five dimensional operator, $\frac{\epsilon}{\Lambda}SF^{\mu\nu}F_{\mu\nu}$
can induce DM decay into monochromatic gamma rays. But for $m_{DM}\sim 100$ GeV and 
cut-off at EW scale, $\Lambda \sim v$,
the DM lifetime again sets $\epsilon \sim 10^{-25}-10^{-27}$.} Therefore, it's natural to adopt a 
$Z_2$ symmetry under which all SM particles
are $Z_2$ even and extra scalars are $Z_2$ odd such a way that the new couplings don't arise in the
Lagrangian which will lead to the decay of dark matter. Also this $Z_2$ symmetry becomes 
accidental for representations $J\geq 2$ if we only allow renormalizable terms in the Lagrangian. 
Moreover, our study is performed in a
region of parameter space where inert scalar multiplets do not develop
any vev both in zero and finite temperature.\footnote{One can consider 
a scenario where inert multiplet can have non-zero
vacuum expectation value at some finite temperature but it relaxes to zero as the temperature lowers down.
Such scenario has been explored for 
doublet \cite{Ginzburg:2010wa} and real triplet \cite{Patel:2012pi}.}

Denoting scalar multiplet as $Q$, and the SM Higgs as $\Phi$, the most
general Higgs-scalar multiplet potential , symmetric under $Z_2$, can
be written in the following form, 
\begin{eqnarray} 
  \label{potq}
  V_0(\Phi,Q)&=&- \mu^2 \Phi^\dagger \Phi + M_Q^2 Q^\dagger Q +
  \lambda_1 (\Phi^\dagger \Phi)^2 + \lambda_2 (Q^\dagger Q)^2+
  \lambda_3 |Q^\dagger T^a Q|^2\nonumber\\ 
  &+&\alpha \Phi^\dagger \Phi Q^\dagger Q+\beta \Phi^\dagger
  \tau^a\Phi Q^\dagger T^a Q+\gamma[ (\Phi^T\epsilon \tau^a\Phi) (Q^T
    C T^a Q)^\dagger+h.c] 
\end{eqnarray}
Here, $\tau^a$ and $T^{a}$ are the $SU(2)$ generators in fundamental
and Q's representation respectively. $C$ is an antisymmetric matrix
analogous to charge conjugation matrix defined as, 
\begin{equation}
  C T^{a} C^{-1}=-T^{aT}
\end{equation} 
$C$, being antisymmetric matrix, can only be defined for even 
dimensional space, i.e only for half integer representation. If the
isospin of the reps. is $j$ then $C$ is $(2j+1)\times (2j+1)$
dimensional matrix (an explicit form of generators and $C$ matrix for 
the quartet are given in appendix (\ref{apa})). The generators are 
normalized in such a way, so  
that they satisfy, for fundamental representation, $Tr[\tau^a 
  \tau^b]=\frac{1}{2} \delta^{ab}$ and for other representations,
$Tr(T^{a} T^{b})=D_{2}(Q)\delta^{ab}$. Also $T^a T^a=C_{2}(Q)$. Here,
$D_{2}(Q)$ and $C_{2}(Q)$ are Dynkin index and second Casimir
invariant for $Q$'s representation. The explicit form of generators, 
$\epsilon$ and $C$ matrix are given in the appendix. Notice that,
$\gamma$ term is only allowed for representation with
$(J,Y)=(\frac{n}{2},1)$.

For the doublet, real triplet, complex
triplet, and the quartet the scalar multiplet $Q$ is respectively 
\begin{equation} 
  \begin{pmatrix}
    C^{+}\\
    D^{0}\equiv \frac{1}{\sqrt{2}}(S+i A)\\
  \end{pmatrix}
  \,,
  \begin{pmatrix} 
    \Delta^{+}\\
    \Delta^{0}\\
    \Delta^{-}
  \end{pmatrix}_{(Y=0)}
  \,,
  \begin{pmatrix}
    \Delta^{++}\\
    \Delta^{+}\\
    \Delta^{0}\equiv\frac{1}{\sqrt{2}}(S+i A)
  \end{pmatrix}_{(Y=2)}
  \,, \textrm{ and }
  \begin{pmatrix}
    Q^{++}\\
    Q^{+}\\
    Q^{0}\equiv \frac{1}{\sqrt{2}}(S+iA)\\
    Q^{'-}
  \end{pmatrix}.
	\label{exprep}
 \end{equation}

In general, for the half-integer representation with
$(J,Y)=(\frac{n}{2},1)$ and the Integer representation with
$(J,Y)=(n,Y = 0 \textrm{ or } \pm 2 T_3)$, the scalar multiplets with 
component fields denoted as $\Delta^{(Q)}$, where $Q$ is the electric
charge, are respectively 
\begin{equation}
  \label{hr}
  \bf{Q_{\frac{n}{2}}}=\begin{pmatrix}
    \Delta^{(\frac{n+1}{2})}\\
    ...\\
    \Delta^{(0)}\equiv\frac{1}{\sqrt{2}}(S+i\, A)\\
    ...\\
    \Delta^{(-\frac{n-1}{2})}\\
  \end{pmatrix}
\, \textrm{ and } \,
  \bf{Q_{n}}= \begin{pmatrix}
    \Delta^{(n)}\\
    ...\\
    \Delta^{(0)}\\
    ...\\
    \Delta^{(-n)}\\
  \end{pmatrix}_{Y=0}.
\end{equation}
For the former representation every component represents a unique field
while for the latter there is a redundancy $\Delta^{(-n)} =
(\Delta^{(n)})^{*}$ except the $Y\neq 0$ case for which the component
are also unique.

\subsection{Inert Triplet and Quartet Mass spectra}

{\bf Mass spectra: Half Integer Representation with $Y=1$}

We now sketch the general form of mass spectrum for the multiplet. As
$Y=1$, $T_3$ value of the neutral component of the multiplet has to be
$T_3=-\frac{1}{2}$. Now for Higgs vacuum expectation value, $\langle
\Phi\rangle=(0,\frac{v}{\sqrt{2}})^{T}$, the term $\langle\Phi^\dagger
\rangle\tau^3\langle\Phi\rangle$ gives $-\frac{v^2}{4}$. So masses for
the neutral components, $S$ and $A$ are respectively 
\begin{equation}  \label{nc}
m_{S}^2 = M_Q^2+\frac{1}{2}(\alpha+\frac{1}{4}\beta+p(-1)^{p+1}\gamma)
v^2 \, \textrm{ and }
m_{A}^2=M_Q^2+\frac{1}{2}(\alpha+\frac{1}{4}\beta-p(-1)^{p+1}\gamma)
v^2.
\end{equation}
Here, $p=\frac{1}{2}\text{Dim}(\frac{n}{2})=1,2,...$ comes from
$2p\times2p$ $C$ matrix. For the charged component, with $T_3=j$, we
have, 
\begin{equation}
  m_{(j)}^2=M_Q^2+\frac{1}{2}(\alpha-\frac{1}{2} j\beta) v^2.
\end{equation}

Now because of the $\gamma$ term, there will be mixing between
components carrying same amount of charge. So to write down the mixing
matrix, we have considered the ordering as follows. A component of the
multiplet is denoted as $|J,T_3\rangle$. Components below the neutral
component $|\frac{n}{2},-\frac{1}{2}\rangle$ are denoted with
$|\frac{n}{2},-\frac{1}{2}-m\rangle$ where, $m=1,2,....,\frac{n-1}{2}$
and corresponding charge is $Q=-m$. The piece
$\langle\Phi\rangle^T\epsilon \tau^a\langle\Phi\rangle$ gives
$\frac{v^2}{2\sqrt{2}}$. Therefore, the mixing term between between
components with charge $|Q|=m$ is, $(-1)^{m+1}\frac{\gamma
  v^2}{4}\sqrt{(n+2m+1)(n-2m+1)}$. In the ordering,
$(\Delta_{(\frac{1}{2})}^{+},\Delta_{(-\frac{3}{2})}^{+})$, with
$m=1$, the mass matrix becomes, 
\begin{equation}
  \label{sc1}
  M^2_{+}=\begin{pmatrix}
  m^2_{(\frac{1}{2})}&\,\, \frac{\gamma v^2}{4}\sqrt{(n+3)(n-1)}\\
  \\
  \frac{\gamma v^2}{4}\sqrt{(n+3)(n-1)}&\,\, m^2_{(-\frac{3}{2})}
  \end{pmatrix}
\end{equation}
And with $m=2$ and
$(\Delta_{(\frac{3}{2})}^{++},\Delta_{(-\frac{5}{2})}^{++})$ we have, 
\begin{equation}
  \label{sc2}
  M^2_{++}=\begin{pmatrix}
  m^2_{(\frac{3}{2})}&\,\, -\frac{\gamma v^2}{4}\sqrt{(n+5)(n-3)}\\
  \\
  -\frac{\gamma v^2}{4}\sqrt{(n+5)(n-3)}&\,\, m^2_{(-\frac{5}{2})}
  \end{pmatrix}
\end{equation}
and so on, for charges with $m=3,4, \ldots$.
\\

For example, if we consider $S$ of inert doublet to be the DM candidate,
then following our parameterization Eq.(\ref{potq}) and Eq.(\ref{exprep}),
the mass spectrum, where the mass hierarchy becomes apparent, is in the following

\begin{eqnarray}
m^2_{S}&=&M^2_{Q}+\frac{1}{2}(\alpha+\frac{1}{4}\beta+\gamma)v^2\nonumber\\
m^2_{A}&=&m^2_{S}-\gamma v^2\nonumber\\
m^2_{C}&=&m^2_{S}-\frac{1}{2}(\gamma+\frac{1}{2}\beta)v^2
\label{doubetmass}
\end{eqnarray}

{\bf Quartet Model\label{quartetsec}} 
The immediate generalization of doublet case is the $J=3/2$  quartet
case. Therefore we have carefully investigated its relic density,
direct detection, it's connection to strong EWPhT in low mass regime
of DM. Apart from splitting between $S$ and $A$,
$\gamma$ term also mixes two single charged components of the
quartet. According to Eq.(\ref{sc1}), the mass matrix for single
charged fields in $(Q^{+},Q^{'+})$ basis is 
\begin{equation}
  M^2_{+}=\begin{pmatrix}
  M_Q^2+\frac{1}{2}(\alpha-\frac{1}{4}\beta)v^2& \frac{\sqrt{3}}{2}\gamma v^2\\
  \frac{\sqrt{3}}{2}\gamma v^2& M_Q^2+\frac{1}{2}(\alpha+\frac{3}{4}\beta)v^2
  \end{pmatrix}
\end{equation}
Diagonalizing the mass matrix, we have mass eigenstates for single
charged fields, $Q_1^+=Q^{+}\cos\theta +Q^{'+}\sin\theta$,
$Q_2^+=-Q^{+}\sin\theta +Q^{'+}\cos\theta$ with
$\tan2\theta=-\frac{2\sqrt{3}\gamma}{\beta}$. 

Again we consider $S$ to be the dark matter. So masses of the components of the multiplet are 
as follows,

\begin{eqnarray}
\label{qrmass1}
m^2_{S}&=&M^2_{Q}+\frac{1}{2}(\alpha+\frac{1}{4}\beta-2\gamma)v^2\nonumber\\
m^2_{A}&=&m^2_{S}+2\gamma v^2\nonumber\\
m^2_{Q^{++}}&=&m^2_{S}-\frac{1}{2}(\beta-2\gamma)v^2\nonumber\\
m^2_{Q^{+}_{1}(Q^{+}_{2})}&=&m^2_{S}+(\gamma \mp \frac{1}{4}\sqrt{\beta^2+12\gamma^2})v^2
\end{eqnarray}

Because of the mixing between two single charged states, the mass
relation is 
\begin{equation}
  m_{S}^2+m_{A}^2=m_{Q_{1}^{+}}^2+m_{Q_{2}^{+}}^2
\end{equation}

The full stability analysis for the quartet potential is an involved task but for time being,
we can give a partial set of 
necessary stability conditions by taking 
Higgs-scalar two dimensional complex surface $(h, \xi^i,0,0,...)$ in field 
space.\footnote{For general treatment see \cite{Kannike:2012pe}}.

\begin{eqnarray}
  \label{stq}
  \lambda_1>0 \, &,& \lambda_2,\lambda_3>0 \nonumber\\
  \alpha+\frac{1}{4}\beta-2\gamma &>& -2\sqrt{\lambda_1( \lambda_2+\frac{1}{4}\lambda_3)} \,\,\, \text{in $(h,S,0,..)$ surface} \nonumber\\
  \alpha+\frac{1}{4}\beta+2\gamma &>& -2\sqrt{\lambda_1( \lambda_2+\frac{1}{4}\lambda_3)} \,\,\, \text{in $(h,A,0,..)$ surface} \nonumber\\
  \alpha-\frac{3}{4}\beta &>& -2\sqrt{\lambda_1( \lambda_2+\frac{3}{4}\lambda_3)} \,\,\, \text{in $(h,Q^{++},0,..)$ surface}\nonumber\\
  \alpha-\frac{1}{4}\beta &>& -2\sqrt{\lambda_1( \lambda_2+\frac{1}{4}\lambda_3)} \,\,\, \text{in $(h,Q^+,0,..)$ surface} \nonumber\\
  \alpha+\frac{3}{4}\beta &>& -2\sqrt{\lambda_1( \lambda_2+\frac{9}{4}\lambda_3)} \,\,\, \text{in $(h,Q'^-,0,..)$ surface}
\end{eqnarray}

The last two conditions, when expressed in terms of mass eigenstates ${Q^+_1,Q^+_2}$ becomes
\begin{equation} \label{stq1}
  \alpha\mp \frac{1}{4}\beta (1 \mp 2 \cos{2\theta}) \pm
  \sqrt{3}\gamma\sin{2\theta} > -\sqrt{\lambda_1
    \lambda_2+\frac{1}{4}\lambda_1\lambda_3(3 \mp
    4\cos{2\theta}+2\cos{4\theta})}
\end{equation}
in $(h,Q_{1,2}^+)$ surface.
\\

{\bf Mass spectra: Integer Representation with $Y=0$ and $Y\neq 0$}
For integer representation, $\gamma$ term is not allowed. Therefore, there will not be 
any mass splitting between real and imaginary part of the neutral component. 
Moreover, for $Y=0$ or real representation, the term $Q^\dagger T^3 Q$ is zero. So tree 
level mass spectrum is degenerate and is given by,
\begin{equation}
  m^2_{\Delta}=M^2_Q+\frac{1}{2} \alpha v^2.
\end{equation}
On the other hand, if $Y\neq 0$, there will be mass splitting due to
$\beta$ term. Also one can choose $Y$ as, from $-n$ to $n$, to set one
component to be neutral. If $T_3=j$, the mass is given by, 
\begin{equation}
  m^2_{(j)}=M^2_Q+\frac{1}{2}(\alpha-\frac{1}{2}\beta j)v^2
\end{equation}

{\bf Triplet Model\label{tripletsec}}

A well motivated representative of this
class is the Triplet with $Y=2$ because of it's role in explaining the
smallness of neutrino mass in Type-II seesaw mechanism
\cite{typeII}. In this case, $\Delta$ is odd under $Z_2$ symmetry, therefore the neutral
component will not acquire any vacuum expectation
value. For real triplet, the term, $\Delta^\dagger T_T^3 \Delta$ is zero,
so the mass spectrum will be degenerate and given as, 
\begin{equation}
  m^2_{\Delta^{+}}=m^2_{\Delta^{0}}=M_\Delta^2+\frac{1}{2}\alpha v^2.
\end{equation}

On the other hand, for complex case, the mass spectrum is, 
\begin{eqnarray}
  \label{mt}
  m_{S}^2=m_{A}^2&=&M_\Delta^2+\frac{1}{2}(\alpha+\frac{1}{2} \beta) v^2\\
  m_{\Delta^{+}}^2&=&M_\Delta^2+\frac{1}{2}\alpha v^2\\
  m_{\Delta^{++}}^2 &=&M_\Delta^2+\frac{1}{2}(\alpha-\frac{1}{2} \beta) v^2 
\end{eqnarray}
There is a relation between masses:
\begin{equation}
  m_{\Delta^{++}}^2-m_{\Delta^{+}}^2=m_{\Delta^{+}}^2-m_{\Delta^{0}}^2=\frac{1}{4}\beta v^2
\end{equation}

And the stability conditions are,
\begin{eqnarray}
  \label{stt}
  \lambda_1>0 \, &,& \lambda_2,\lambda_3>0\nonumber\\
  2\alpha+\beta &>& -2\sqrt{\lambda_1( \lambda_2+\lambda_3)} \,\,\, \text{in $(h,\Delta^0,0,..)$ surface} \nonumber\\
  \alpha &>& -2\sqrt{\lambda_1 \lambda_2} \,\,\, \text{in $(h,\Delta^+,0,..)$ surface}\\
  2\alpha-\beta &>& -2\sqrt{\lambda_1( \lambda_2+\lambda_3)} \,\,\, \text{in $(h,\Delta^{++},0,..)$ surface}
\end{eqnarray}

Because of the absence of $\gamma$ term as in half integer
representation, there is no mass splitting between $S$ and $A$
of the neutral component $\Delta^{0}$. $S$ and $A$ have vector interaction with $Z$ boson that produces
spin-independent elastic cross section which is already $8-9$ orders
of magnitude above the CDMS direct detection bound \cite{Akerib:2005kh} and
for this reason, complex triplet or any larger integer multiplet with non zero 
hypercharge is not going to satisfy direct detection bound as
one will require a non-zero splitting between $S$
and $A$ larger than the kinetic energy of DM in our galactic halo to make
$Z-S-A$ process kinematically forbidden. Therefore, only
neutral component of the real triplet can be a plausible DM candidate. In
fact real triplet DM has already been studied in \cite{Hambye:2009pw,
  Cirelli:2005uq, Cirelli:2007xd, FileviezPerez:2008bj}. In
\cite{Cirelli:2005uq, Cirelli:2007xd} it was shown that observed relic
density can be accounted for DM if the mass of DM lies about $2.5$
TeV. Being degenerate, one can easily see that, such particle would
decouple from electroweak plasma and had no significant effect on
EWPhT. But the question of having a strong first order EWPhT in 
complex triplet is relevant for cosmic evolution of the universe. 
Therefore we only focused on EWPhT in complex triplet case in a later section.

\subsection{Perturbativity and EW Physics Constraints on the Size of Multiplets}

{\bf Perturbativity and Renormalization Group Equations\label{rgeq}}

What is the largest possible inert multiplet allowed to be added to
the standard model? One possible bound comes from the beta function 
of $SU(2)$ gauge coupling in presence of large scalar multiplets because the
addition of such large multiplet not only halts the asymptotic freedom of 
non-Abelian gauge couplings but also lowers the scale of landau pole with its size.

\begin{figure}[h!]
  \centerline{\includegraphics[width=8cm]{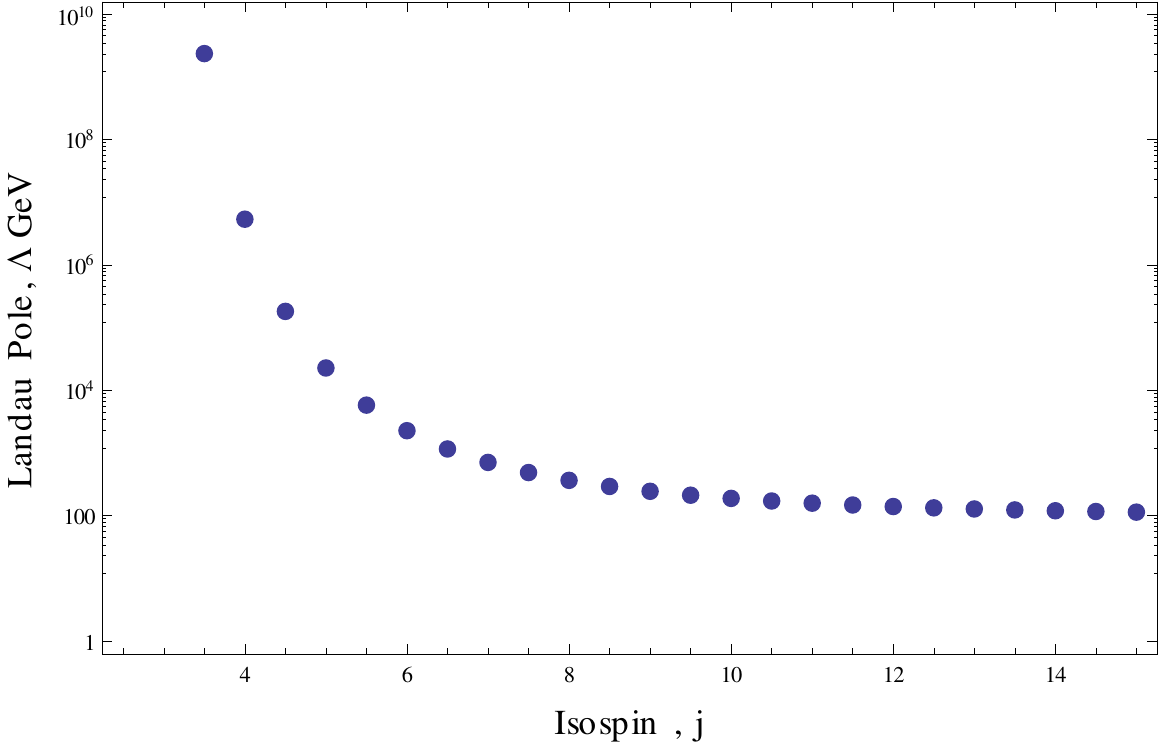}}
  \caption{Landau pole with different scalar multiplet.} 
  \label{rg} 
\end{figure}

The one-loop beta function of $SU(2)$ gauge coupling for Standard Model 
solely augmented by a scalar multiplet of isospin $J$ is $\beta(g)=\frac{g^3}{16
  \pi^2}(-\frac{19}{6}+\frac{1}{9}J(J+1)(2J+1))$. It can be seen that, $\beta(g)$ 
remains negative only for $J\leq
\frac{3}{2}$. For, $J\geq 2$, it becomes positive and hits the landau
pole as shown in Fig.(\ref{rg}). 
For instance adding a scalar multiplet with isospin $J\geq 5$ will bring
the Landau pole of $SU(2)$ gauge coupling at
$\Lambda\leq 10$ TeV and for $J\geq 10$, its even smaller, $\Lambda\leq 180$ GeV.
Therefore, perturbativity of gauge coupling at the TeV scale sets a upper bound on the 
size of the multiplet to be $J\leq 5$.

Another bound on the size of the multiplet, charged under SM gauge
group, is set by perturbative unitarity of tree-level scattering
amplitude. In \cite{Hally:2012pu, Earl:2013jsa}, the $2\rightarrow 2$ scattering amplitudes for
scalar pair annihilations into electroweak gauge bosons have been
computed and by requiring zeroth partial wave amplitude satisfying the
unitarity bound, it was shown that maximum allowed complex $SU(2)_L$
multiplet would have isospin $J\leq 7/2$ and real multiplet would have
$J\leq 4$.

\begin{figure}[h!]
  \centerline{\includegraphics[width=10cm]{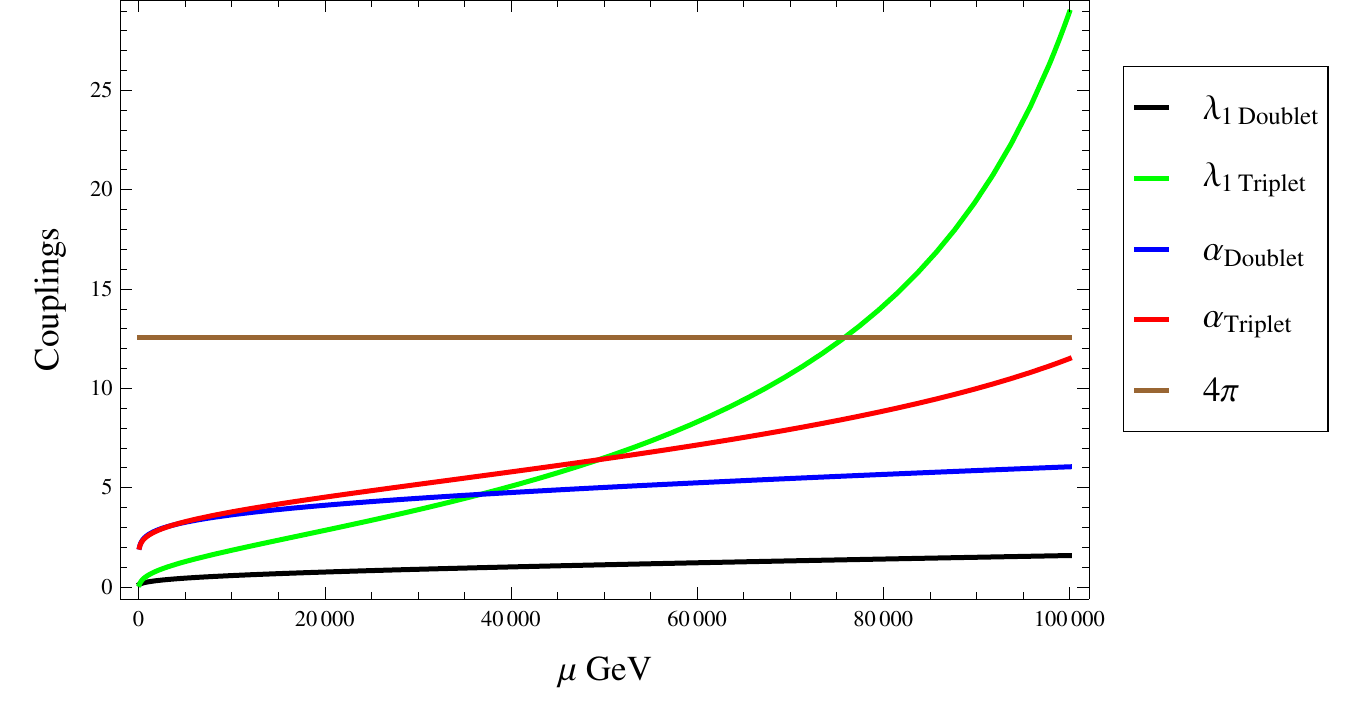}}
  \caption{The running of Higgs quartic coupling $\lambda_{1}$ and coupling $\alpha$. 
The initial values of the couplings at energy $\mu=100$ GeV 
are $(\lambda_{1},\lambda_{2},\lambda_{3},\alpha,\beta,\gamma)=(0.13,0.1,0.1,2,0.1,0.1)$ for 
both cases. Here we can see that large values of $\alpha$ which also drives strong EWPhT, 
as we will see later, drives Higgs coupling to run much faster in triplet than the doublet case.} 
  \label{rgcompare} 
\end{figure}

In addition, a check with 1-loop beta functions, given in appendix.(\ref{rgeqn}), for 
the triplet \cite{Schmidt:2007nq} compared to 
the doublet \cite{Barbieri:2006dq} shows in Fig. (\ref{rgcompare}) that on increasing 
the size of the multiplet, the scalar couplings will run faster compared to the smaller 
representation and will become non-perturbative much faster if one starts with large scalar
coupling at EW scale.

{\bf Electroweak Precision Observable\label{ewpo}}

A straightforward way to observe the indirect effect of new physics is
in the modification of vacuum polarization graph of $W^{\pm}$ and
$Z^0$ boson and one convenient way to parameterize these 'Oblique
correction' is through $S$, $T$ and $U$ parameters
\cite{Kennedy:1988sn, Peskin:1991sw} (analysis of one-loop correction
in SM was first done in \cite{Passarino:1978jh}); plus V, W and X parameters
\cite{Maksymyk:1993zm} if new physics is at scale comparable to the EW
scale. Oblique corrections are dominant over other 'non-oblique' corrections 
(vertex diagram and box diagram with SM fermions as external states)
because all the particles charged under SM group will couple to gauge
bosons but usually only one or two particles in a theory will couple
to specific fermion species. 

In the present case we have only considered the quartet and because of absence 
of coupling between quartet fields and SM fermions, the only dominant effects will come
from oblique corrections. $T$ parameter, measuring the 
shift of $\rho=\frac{M_{W}}{M_{Z}\cos\theta_{w}}$ from 
SM value due to the radiative correction by new particles, is 
\begin{equation}
  \alpha T=\frac{\Pi_{WW}(0)}{M^2_{W}}-\frac{\Pi_{ZZ}(0)}{M^2_{Z}}
\end{equation}
where, $\Pi_{WW}(0)$ and $\Pi_{ZZ}(0)$ vacuum polarization graph of $W$
and $Z$ bosons evaluated at external momentum, $p^2=0$.

And quartet contribution to T parameter,
\begin{eqnarray}
\label{tp} 
  \frac{16\pi^2\alpha}{g^2} T &=&\frac{1}{ M^2_{W}}(\frac{3}{2} \cos^2\theta \,
  F(m_{Q^{++}},m_{Q_{1}^{+}})+\frac{3}{2} \sin^2\theta\,
  F(m_{Q^{++}},m_{Q_{2}^{+}})\nonumber\\ 
  &+&\frac{1}{4}(\sqrt{3}\sin\theta+2\cos\theta)^2\,
  F(m_{Q_{1}^{+}},m_{S})+\frac{1}{4}(\sqrt{3}\sin\theta-2\cos\theta)^2\,
  F(m_{Q_{1}^{+}},m_{A})\nonumber\\ 
  &+&\frac{1}{4}(\sqrt{3}\cos\theta-2\sin\theta)^2\,
  F(m_{Q_{2}^{+}},m_{S})+\frac{1}{4}(\sqrt{3}\cos\theta+2\sin\theta)^2\,
  F(m_{Q_{2}^{+}},m_{A}))\nonumber\\ 
  &-&\frac{1}{4 c^2_{w} M^2_{Z}}(F(m_{S},m_{A})+\sin^2 2\theta \,
  F(m_{Q_{1}^{+}},m_{Q_{2}^{+}})
\end{eqnarray}

here,
\begin{equation}
  F(m_1,m_2)=\frac{m^2_1+m^2_2}{2}-\frac{m_1^2
    m_2^2}{m_1^2-m_2^2}\ln\frac{m_1^2}{m_2^2} 
\end{equation}

In addition, S parameter for quartet
multiplet is, 
\begin{eqnarray}
  \label{sp}
  S &=&\frac{s_{w}^2}{8\pi}\int_0^1 dx\,
   x(x-1)[\ln\frac{m^2_{Q_{1}^{+}}m^2_{Q_{2}^{+}}}{m^4_{Q^{++}}}+2\cos2\theta\ln\frac{m^2_{Q_{2}^{+}}}{m^2_{Q_{1}^{+}}}\nonumber\\ 
      &+&\ln\frac{(1-x)m^{2}_{S} + x m^{2}_{A}}{m^{2}_{Q^{++}}}+\sin^2
      2\theta\ln\frac{[(1-x)m^2_{Q_{1}^{+}}+xm^2_{Q_{2}^{+}}]^2}{m^2_{Q_{1}^{+}}m^2_{Q_{2}^{+}}}] 
\end{eqnarray}

The best fit values of S and T
parameter with ($U=0$)\footnote{The contribution to $U$ parameter by scalar multiplet with only gauge 
interactions considered in our case, will be smaller
compared to $T$ parameter by a factor $(M_{W}/M_{S})^2$, where $M_{S}$ is the 
leading scalar mass of theory \cite{Barbieri:2004qk}}  is \cite{Beringer:1900zz} 
\begin{equation}
  S=0.04\pm0.09\,\,\,\text{and}\,\,\,T=0.07\pm0.08
\end{equation}

Therefore, one can put constraints on S and T parameter by
comparing the theoretical predictions with well measured
experimental values of observables.

\section{Electroweak Phase Transition (EWPhT)\label{ewpht}}
\subsection{Finite Temperature Effective Potential}
To study the impact of large scalar multiplets on the electroweak
phase transition, we have used the standard techniques of finite
temperature field theory \cite{Kirzhnits:1976ts, Weinberg:1974hy,
  Dolan:1973qd, Mohapatra:1979qt, Mohapatra:1979vr} (for a quick review, also see
\cite{Quiros:1999jp}). If there are multiple classical background
fields  $\phi_i$ , which act as order parameters of the thermodynamic
system, the total one-loop effective potential at finite temperature
is, 
\begin{equation}
  \label{eftp}
  V_{eff}(\phi_i,T)=V_{0}(\phi_i)+V_{CW}(\phi_i)+V_{T}(\phi_i,T)
\end{equation}
Here, $V_{0}$, $V_{CW}$ and $V_{T}$ are tree-level, 1-loop
Coleman-Weinberg and finite temperature potential respectively. The
daisy resummed \cite{Parwani:1991gq, Arnold:1992rz} finite temperature
potential is, 
\begin{equation}
  \label{finited}
  V_{T}=\sum_{B(F)}(\pm) g_{B(F)}\frac{T^4}{2 \pi^2}\int_0^\infty
  dxx^2\ln(1\mp e^{-\sqrt{x^2+m_{B(F)}^2(\phi_i,T)/T^2}}) 
\end{equation}
Here, $g_B$ and $g_F$ are bosonic and fermionic degrees of freedom and
$\pm$ correspond to boson and fermion respectively. Thermal mass
correction determined with respect to background fields $\phi_{j}$ is, 
\begin{equation}
  \label{daisy}
  m^2_{i}(\phi_{j})\rightarrow
  m^2_{i}(\phi_{j},T)=m^2_{i}(\phi_{j})+\Pi_{i}(T) 
\end{equation}
where $\Pi_{i}(T)$ is the thermal self energy correction (Debye
correction) and at the high temperature limit, it is of the form $T^2$
times coupling constants.  $\Pi_{i}(T)$ measures how much particles
are screened by thermal plasma from the classical background field
$\phi$ (just like the Debye screening) and large screening reduces the
strength of phase transition. In other words, it is the amplitude for
the external particle (sourced by classical field) to forward scatter
off from a real physical particle present in the thermal bath
\cite{Arnold:1994bp}. 

For numerical convenience, in subsequent studies, we have used the
following form of the effective potential in high temperature
approximation Appendix (\ref{ahigheff}), 
\begin{eqnarray}
  \label{heff}
  V_{eff}&=&V_0+V_{CW}+\sum_{B}g_B[\frac{m^2_B(\phi,T)
      T^2}{24}-\frac{T}{12
      \pi}[m^2_B(\phi,T)]^{\frac{3}{2}}\\ \nonumber 
    &+&\frac{m^4_B(\phi,T)}{64\pi^2}\ln\frac{m^2_B(\phi,T)}{A_b
      T^2}]+\sum_{F}g_F[\frac{m_F^2(\phi)
      T^2}{48}-\frac{m^2_F(\phi)}{64\pi^2}\ln\frac{m^2_F(\phi)}{A_f
      T^2}] 
\end{eqnarray}
It was shown in \cite{Anderson:1991zb} that high T approximation
agrees with exact potential better than $5\%$ for $m/T<1.6(2.2)$ for
fermions (bosons). So unless great accuracy is required, one can use
Eq.(\ref{heff}) to explore the thermodynamic properties of system. 

There will be some region of parameter space, for example, the
Goldstone modes, where the effective potential will become imaginary
due to the non-analytic cubic terms $(m^2(\phi,T))^{\frac{3}{2}}$ and
the log terms. It doesn't signal the breakdown of perturbative
calculation, instead, as it was shown in \cite{Weinberg:1987vp,
  Boyanovsky:1996dc} that, imaginary part signals the instability of
the homogeneous zero modes. Moreover, as mentioned in
\cite{Weinberg:1973am} that the imaginary part vanishes when effective
potential is calculated to all orders but at any finite order, it can
be present. Therefore, in calculation we are only concerned with the
real part of the potential. To ameliorate the gauge dependence of
finite temperature effective potential, gauge invariant prescriptions
have been developed \cite{Nielsen:1975fs, Buchmuller:1994vy,
  Boyanovsky:1996dc} and also recently in
\cite{Patel:2011th}. Moreover, in \cite{Boyanovsky:1996dc}, it was
shown that landau gauge ($\xi=0$) is better in capturing the
thermodynamic properties by comparing them with those determined with
gauge invariant Hamiltonian formalism. So in this paper we have chosen
to follow the landau gauge to carry out our numerical calculation.

The main motivation for us to explore the phase transition
qualitatively and it's already apparent from above discussion that
perturbative techniques can at best captures approximate nature of
finite temperature phenomena. For quantitative accuracy one have to
use lattice methods.  

\subsection{EWPhT with Inert Triplet and Quartet Representations}
The nature of electroweak phase transition is a cross over for Higgs
boson with mass about $125.5$ GeV (for recent lattice study,
\cite{Laine:2012jy}). Therefore to achieve strong first order phase
transition, one must extend the scalar sector of the theory. Consider
an inert multiplet $Q$ with isospin, $j$ and hypercharge $Y=0$ and the
parameter space is chosen in such a way that inert multiplet does not
obtain any VEV at all temperatures. So, $\langle
\Phi\rangle=(0,\frac{\phi}{\sqrt{2}})^{T}$ and $\langle
Q\rangle=0$. Because of zero VEV at all temperatures, the only
classical background field is that of Higgs doublet and therefore the
sphaleron configuration is exactly like that of standard model. For
this reason, in this case, the first order phase transition is
determined by the condition $\phi_c/T_c \geq 1$ (for a quick review,
\cite{Quiros:1999jp}). 

The thermal masses for the component fields of the Higgs doublet
$\Phi$ and real multiplet $Q$ are 
\begin{eqnarray}
  \label{rhiggs}
  m^2_{h}(\phi, T)&=&-\mu^2+3\lambda_{1}
  \phi^2+a(j) \frac{T^2}{12}\\ \nonumber 
  m^2_{G^{\pm}}(\phi, T)=m^2_{G^{0}}(\phi,T)&=&-\mu^2+\lambda_{1}
  \phi^2+a(j) \frac{T^2}{12} 
\end{eqnarray}
and due to degenerate mass spectrum,
\begin{equation}
  \label{realm}
  m^2_{i}(\phi,T)=M^2+\frac{1}{2}\alpha\phi^2+b(j)\frac{T^2}{12} 
\end{equation}

Temperature coefficients are,
\begin{eqnarray}
  \label{tfactor}
  a(j)&=&6\lambda_{1}+\frac{1}{2}(2j+1)\alpha+\frac{9}{4}g^2+\frac{3}{4}g'^2+3y_t^2\\ 
  b(j)&=&(2j+3)\lambda_2+2\alpha+3j(j+1)g^2 
\end{eqnarray}

Here one can see that from 1st and 3rd term of $b(j)$ that larger
representation receives relatively large thermal corrections due to
the scalar loops and gauge boson loops. These coefficients capture how
much particles are screened by the plasma from the classical field
which determines the strength of the transition. The larger the
coefficients are, the weaker the transition will become because those
particles are effectively decoupled from the plasma. This plasma
screening effect on the nature of phase transition for complex singlet
was already studied in \cite{Espinosa:1993bs}. In the following, we
showed similar effect for real multiplet as it captures the essential
features that depend on size of the multiplet. Generalization to
complex odd dimensional or even dimensional scalar multiplet is
straightforward. 

Using Eq.(\ref{heff}) and neglecting the terms coming from
CW-corrections and log terms we have, 
\begin{equation}
  V_{T}(\phi, T)=A(T) \phi^2+ B(T) \phi^4+C(T)[\phi^2+K^2(T)]^{\frac{3}{2}}
\label{htpot}
\end{equation}
where,
\begin{eqnarray}
  A(T)&=&-\frac{1}{2}\mu^2+a(j)\frac{T^2}{12}\\ \nonumber 
  B(T)&=&\frac{1}{4}\lambda_{1}\\ \nonumber 
  C(T)&=& -(2j+1)\frac{T}{12
    \pi}(\frac{\alpha}{2})^{\frac{3}{2}}\\ \nonumber 
  K^2(T)&=& \frac{2}{\alpha}(M^2+b(j)\frac{T^2}{12})
\label{htpotc} 
\end{eqnarray}

When the universe is at very high temperature, it is in the symmetric
vacuum $\phi=0$ but when the universe cools down, there can be two
characteristic temperatures: $T_1$ and $T_2$. For temperature,
$T<T_2$, the origin is the maximum and there is only one global
minimum at $\phi\neq 0$ that evolves towards zero temperature
minimum. For $T>T_2$, the origin is a minimum and there is also a
maximum at $\phi_{-}(T)$ and another minimum at $\phi_{+}(T)$ given
by, 

\begin{equation}
  \phi^2_{\pm}(T)=\frac{1}{32 B^2}[9 C^2- 16 AB\pm 3|C|\sqrt{9 C^2+32(2 B^2 K^2- AB}]
\end{equation}
The second order transition temperature $T_2$ is determined by the condition,
\begin{equation}
  4 A^2-9 C^2 K^2=0
\end{equation}
And the first order transition temperature, $T_1$ where
$V_{T}(\phi_{c}(T_{1}), T_{1})=V_{T}(0, T_{1})$ and
$\phi_{+}(T_1)=\phi_{-}(T_1)$ sets the condition, 
\begin{equation}
  9 C^2+32(2 B^2 K^2-AB)=0
\end{equation}
From two conditions, $T_1$ and $T_2$ are determined as,
\begin{equation}
  \label{fot}
  T_1^2=\frac{4\lambda_{1} \mu^2+\frac{8
      M^2+\lambda_{1}^2}{\alpha}}{\frac{1}{3}a(j)\lambda_{1}-\frac{2b(j)\lambda_{1}^2}{3\alpha}-(2j+1)^2\frac{\alpha^3}{128\pi^2}}  
\end{equation}
and 
\begin{equation}
  \label{sot}
  T_2^2=\frac{1}{2 D}(E+\sqrt{E^2-4 D \mu^4}) 
\end{equation}
with
\begin{eqnarray}
  D&=&\frac{a(j)^2}{144}-(2j+1)^2\frac{b(j)\alpha^2}{768\pi^2}\\ \nonumber 
  E&=&\frac{1}{6}a(j)\mu^2+(2j+1)^2\frac{M^2\alpha^2}{64\pi^2}
\end{eqnarray}

The nature of the transition depends on the relation between $T_1$ and
$T_2$. If $T_1>T_2$, the transition is first order and plasma
screening is not so effective. When $T_1<T_2$, the transition is
actually second order due to dominant plasma screening. Actually,
$T_1=T_2$ gives the turn over condition from first to second order
transition and from (\ref{fot}) and (\ref{sot}), we can have a
condition on the parameter space $(\lambda_{2},\alpha,M,j)$. 

\begin{equation}
  \label{ine}
  \frac{3}{256\pi^2} (2j+1)^2 \alpha^4-b(j) \lambda_{1}^2\geq ( a(j)
  \lambda_{1}-\frac{3}{64 \pi^2}(2j+1)^2 \alpha^3)(\frac{M}{v})^2 
\end{equation}

Here the strict inequality implies the region of parameter space where
first order transition persists and equality corresponds to the
turn-over. Here one can see that, this inequality saturates if $M$ and
$b(j)$ becomes large as these two terms control the plasma screening
for the particle. For small values of $\alpha$, we can easily see from
LHS of (\ref{ine}) that the second term increases faster than the
first term due to the quadratic Casimir for gauge boson contribution
and self interacting quartic term in $b(j)$ (Eq.(\ref{tfactor})). Also
if invariant mass term $M$ is large, the RHS will saturates the
inequality much faster. So one can infer that although large
representation will favor the first order transition up to certain
value of $j$ because of more degrees of freedom in the plasma coupling
to the background field, at one point, due to large thermal mass
coming from gauge interaction, plasma screening will be large enough
to cease the first order transition and make it as a second
order. Electroweak phase transition with scalar singlet,  doublet\footnote{Inert doublet can be 
considered a special case of two Higgs doublet model. EWPhT in two Higgs doublet model is 
also studied extensively \cite{Bochkarev:1990gb, Turok:1991uc, Cline:1995dg, Cline:1996mga, 
Cline:2011mm, Shu:2013uua} and references therein} and
real triplet \cite{Patel:2012pi} have been studied extensively. Therefore,
in the following sections, we have focused on immediate extension
i.e. complex triplet and quartet representation to study the nature of
phase transition.

\paragraph{Complex Triplet}
The thermal mass for the Higgs and Goldstone fields are
\begin{eqnarray}
  m_{h}^2(\phi,T)=-\mu^2+3\lambda_{1} \phi^2+a\frac{T^2}{12}\\
  m^2_{G^{\pm}}=m^2_{G^{0}}=-\mu^2+\lambda_{1} \phi^2 +a \frac{T^2}{12}
\end{eqnarray}
And thermal masses for the component fields of the triplet are
\begin{eqnarray}
  \label{mtt}
  m_{S}^2(\phi,T)=m_{A}^2(\phi,T)&=&M_\Delta^2+\frac{1}{2}(\alpha+\frac{1}{2}
  \beta) \phi^2+b \frac{T^2}{12}\\ 
  m_{\Delta^{+}}^2(\phi,T)&=&M_\Delta^2+\frac{1}{2}\alpha
  \phi^2+b\frac{T^2}{12}\\ 
  m_{\Delta^{++}}^2(\phi,T)
  &=&M_\Delta^2+\frac{1}{2}(\alpha-\frac{1}{2} \beta)
  \phi^2+b\frac{T^2}{12} 
\end{eqnarray}
Here the thermal coefficients $a$ and $b$ are,
\begin{eqnarray}
  a&=&6\lambda_{1}+3\alpha+\frac{9}{4}g^2+\frac{3}{4}g'^2+3 y_{t}^2\\ 
  b&=&8\lambda_2+6\lambda_3+2\alpha+6 g^2+3g'^2 
\end{eqnarray}
So the effective potential is
\begin{eqnarray}
  V_{eff}(\phi,T)&=&\frac{1}{2}(-\mu^2+a\frac{T^2}{12})\phi^2+\frac{\lambda_{1}}{4}\phi^4+\sum_{i}
  (\pm)g_i \frac{m_i(\phi)^4}{64
    \pi^2}[\ln\frac{m^2_{i}(\phi)}{Q^2}-c_{i}]\\ \nonumber 
  &-&\frac{T}{12 \pi}\sum_{B}
  g_{B}[m^2_{B}(\phi,T)]^{3/2}-\sum_{B}\frac{m^4_B(\phi,T)}{64\pi^2}\ln\frac{m^2_B(\phi,T)}{A_b
    T^2}\\ \nonumber 
  &+&12\frac{m^4_t(\phi)}{64\pi^2}\ln\frac{m^2_t(\phi)}{A_f T^2} 
\end{eqnarray}
where bosonic sum is taken over, $h$, $G^{\pm}$, $G^{0}$,
$\Delta^{++}$, $\Delta^{+}$, $\Delta^{0}$, $W^{\pm}$ and $Z$ with
corresponding degrees of freedom, $g_{i}$ are $\{h,G^{\pm}, G^{0}, \Delta^{++},
\Delta^{+}, \Delta^{0}, W^{\pm}, Z,t\}=\{1,2,1,2,2,2,6,3,12\}$. Also $Q$ is the renormalization scale
and in $\bar{MS}$ scheme, $\{c_{S},c_{F},c_{GB}\}=\{3/2,3/2,5/6\}$.

\begin{figure}[h!]
\centerline{\includegraphics[width=9.5cm]{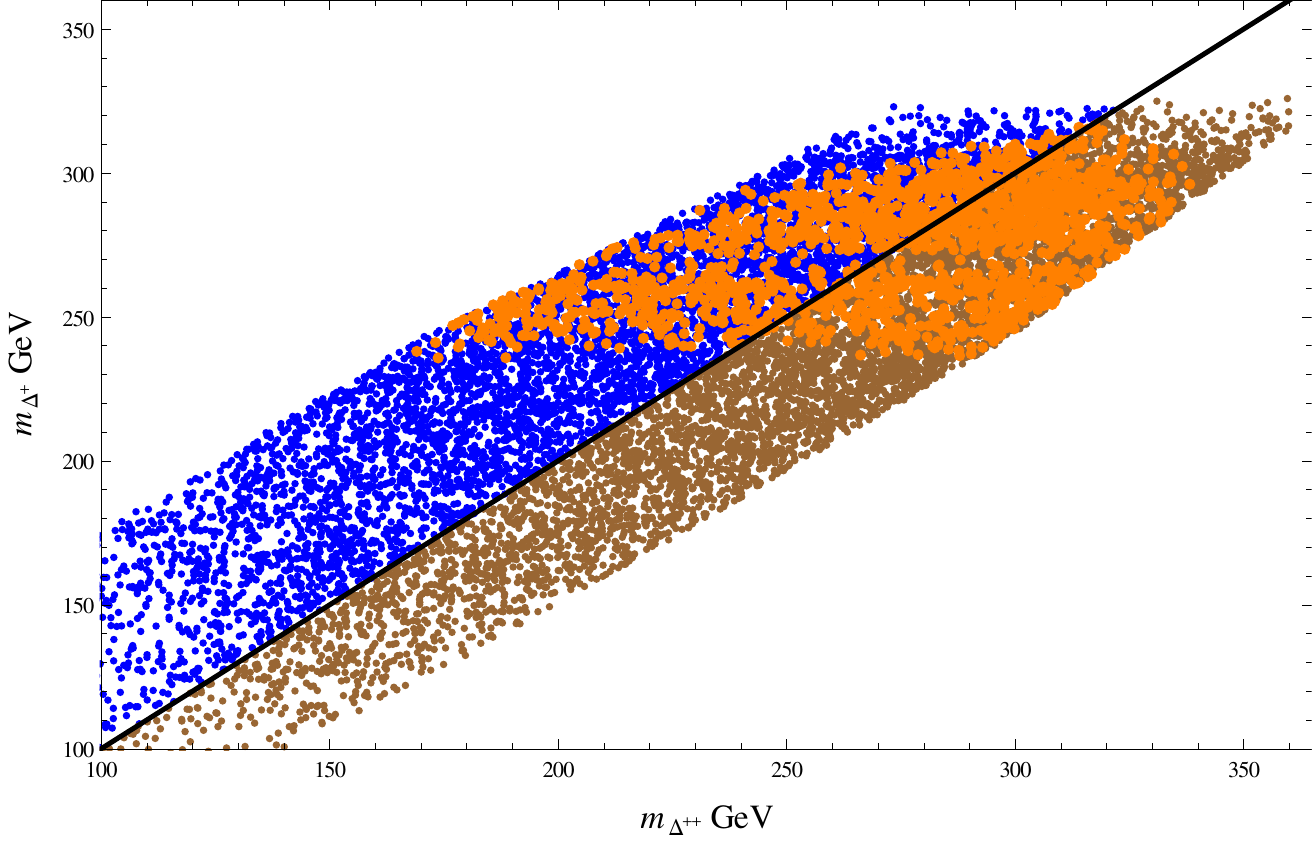}} 
\caption{Correlation between $m_{\Delta^{++}}$ and
  $m_{\Delta^{+}}$. We have scanned the parameter space: $M\in
  (10,150)$ GeV, $\alpha\in (0, 3)$, $\lambda_{2,3}\in (0,0.01)$
  and $|\beta| \in (0,2)$. Here for $\beta
  <0$ (brown points) we have $m_{\Delta^{0}}<m_{\Delta^{+}}<m_{\Delta^{++}}$ and
  strong EWPhT region (orange points) lies for $m_{\Delta^{++}}\sim 250-340$ GeV. For
  $\beta >0$ (blue points) the mass hierarchy is opposite and strong EWPhT region (orange points)
  lies for $m_{\Delta^{++}}\sim 170-315$ GeV. In random scan, for $\beta>0$ and $\beta<0$, 
  out of initial $10^4$ points, $8.54\%$ and $8.79\%$ points which are consistent with stability
  conditions and electroweak precision data (EWPD) showed strong EWPhT respectively. The straight line
  represents $m_{\Delta^{++}}=m_{\Delta^{+}}$.} 
\label{tript}
\end{figure}

In the above analysis, $Z_2$ symmetry is retained but one can introduce 
the term $\mu \Phi^{T}\epsilon \Delta^{\dagger}
\Phi$ which breaks $Z_2$ symmetry softly, which happens in for example,
type-II seesaw model. Such term will induce a
triplet vev, $\langle \Delta^0\rangle=v_{\Delta}$ where
$v_{\Delta}=\frac{\mu v^2}{\sqrt{2}M_{\Delta}^2}$. As indicated in
\cite{Kehayias:2009tn}, such term will modify the Higgs quartic
coupling to $\frac{\lambda_1}{4}\rightarrow
\frac{\lambda_1}{4}-\frac{\mu^2}{2 M_{\Delta}^2}$ which in turn, will
reduce the effective Higgs quartic coupling and enhance the strength
of the transition. The upper bound of $v_{\Delta}$ set by precision measurement 
of $\rho$ parameter, is $2.5-4.6$ GeV \cite{Arhrib:2011uy} and lower bound 
is $10^{-10}$ GeV set by the neutrino mass \cite{Melfo:2011nx}. Now from 
inequality (\ref{ine}), one can see that strong 1st order EWPhT
favors $M_{\Delta} \leq T_c\sim 100-120$ GeV. Therefore $\mu$ will lie
within the range $10^{-11} \leq \mu \leq 1.5$ GeV. Therefore the
correction to the Higgs quartic coupling is about $O(10^{-3})$ so it
is negligible and does not quantitatively change the transition which
is mostly driven by large $\alpha$ coupling of the potential. Also
correction to mass spectrum due to non zero $\mu$ term (therefore,
nonzero $v_{\Delta}$) is $O(\frac{v_{\Delta}^2}{v^2})$ and thus very
small. So EWPhT results obtained in $Z_2$ symmetric triplet case
also holds for softly broken $Z_2$ symmetric model.  

In \cite{ATLAS:2012hi}, for like-sign dilepton final states with
$100\%$ branching ratio at 7 TeV LHC run, the lower limit on mass of
the doubly charged scalar was put as 409 GeV, 398 GeV and 375 GeV for
$e^{\pm}e^{\pm}$, $\mu^{\pm}\mu^{\pm}$ and $e^{\pm}\mu^{\pm}$ final
states. But as pointed out in \cite{Melfo:2011nx} the mass limit
crucially depends on the value of $v_{\Delta}$ and the di-leptonic
decay channel $\Gamma_{\Delta^{++}\rightarrow l_{i}l_{j}}$ is dominant
only when $10^{-10}\leq v_{\Delta}\leq 10^{-5}-10^{-4}$ GeV and when
$v_{\Delta}=10^{-4}-10^{-3}$ it becomes comparable to
$\Gamma_{\Delta^{++}\rightarrow W^{+}W^{+}}$. Also for mass difference
$\Delta M=m_{\Delta^{++}}-m_{\Delta^{+}} \geq 5$ GeV and
$v_{\Delta}\geq 10^{-4}$ GeV, cascade decay is the most dominant decay
channel ($\beta <0$). Therefore when $v_{\Delta}\sim 4\times 10^{-5}$
GeV, di-leptonic branching ratio is around $11\%$ and lower limits on
$m_{\Delta^{++}}$  are 212 GeV ($e^{+}e^{+}$), 216 GeV
($\mu^{+}\mu^{+}$) and 190 GeV ($e^{+}\mu^{+}$)\footnote{Table.1 of
  \cite{ATLAS:2012hi}} which is still compatible with strong EWPhT
region shown in Fig.(\ref{tript}). Moreover, for $v_{\Delta}\geq
10^{-4}$ the limit goes down all the way to $m_{\Delta^{++}}\geq 100$
GeV and thus again compatible with strong EWPhT region.

\paragraph{Quartet Representation}
In case of quartet representation, the thermal mass for the Higgs and Goldstone fields are
\begin{eqnarray}
  m_{h}^2(\phi,T)=-\mu^2+3\lambda_{1} \phi^2+a_{q}\frac{T^2}{12}\\
  m^2_{G^{\pm}}=m^2_{G^{0}}=-\mu^2+\lambda_{1} \phi^2 +a_{q} \frac{T^2}{12}
\end{eqnarray}
And for quartet, the thermal mass for the component fields are
\begin{eqnarray}
  \label{mqt}
  m_{S}^2(\phi,T)&=&M_Q^2+\frac{1}{2}(\alpha+\frac{1}{4}\beta-2\gamma)
  \phi^2+b_{q}\frac{T^2}{12}\\ 
  m_{A}^2(\phi,T)&=&M_Q^2+\frac{1}{2}(\alpha+\frac{1}{4}\beta+2\gamma)
  \phi^2+b_{q}\frac{T^2}{12}\\ 
  m_{Q^{++}}^2(\phi,T) &=& M_Q^2+\frac{1}{2}(\alpha-\frac{3}{4}\beta)
  \phi^2+b_{q}\frac{T^2}{12}\\ 
  m_{Q_{1}^{+}}^2(\phi,T) &=&
  M_Q^2+\frac{1}{2}(\alpha + \frac{1}{4}\beta -
  \frac{1}{2}\sqrt{\beta^2 + 12\gamma^2})\phi^2+b_{q}\frac{T^2}{12}\\  
  m_{Q_{2}^{+}}^2(\phi,T) &=&
  M_Q^2+\frac{1}{2}(\alpha + \frac{1}{4}\beta +
  \frac{1}{2}\sqrt{\beta^2 + 12\gamma^2})\phi^2+b_{q}\frac{T^2}{12}  
\end{eqnarray}
Here the thermal coefficient $a_q$ and $b_q$ are,
\begin{eqnarray}
  a_q&=&6\lambda_{1}+4\alpha+\frac{9}{4}g^2+\frac{3}{4}g'^2+3 y_{t}^2\\
  b_q&=&10\lambda_2+\frac{15}{2}\lambda_3+2\alpha+\frac{45}{4}g^2+\frac{3}{4}g'^2
\end{eqnarray}

Similarly the thermal potential is
\begin{eqnarray}
  V_{eff}(\phi,T)&=&\frac{1}{2}(-\mu^2+a_{q}\frac{T^2}{12})\phi^2+\frac{\lambda_{1}}{4}\phi^4+\sum_{i}
  (\pm)g_{i} \frac{m_i(\phi)^4}{64
    \pi^2}[\ln\frac{m^2_{i}(\phi)}{Q^2}-c_{i}]\\ \nonumber 
  &-&\frac{T}{12 \pi}\sum_{B}
  g_{B}[m^2_{B}(\phi,T)]^{3/2}-\sum_{B}\frac{m^4_B(\phi,T)}{64\pi^2}\ln\frac{m^2_B(\phi,T)}{A_b
    T^2}\\ \nonumber 
  &+&12\frac{m^4_t(\phi)}{64\pi^2}\ln\frac{m^2_t(\phi)}{A_f T^2}
\end{eqnarray}
where bosonic sum is taken over, $h$, $G^{\pm}$, $G^{0}$, $Q^{++}$,
$Q_{1}^{+}$, $Q_{2}^{+}$, $S$, $A$, $W^{\pm}$ and $Z$ with
corresponding degrees of freedom, $\{h,G^{\pm}, G^{0}, Q^{++},
Q_1^{+}, Q_2^{+}, S, A, W^{\pm}, Z,t\}=\{1,2,1,2,2,2,1,1,6,3,12\}$. 
	
\begin{figure}[h!]
\centerline{\includegraphics[width=8.5cm]{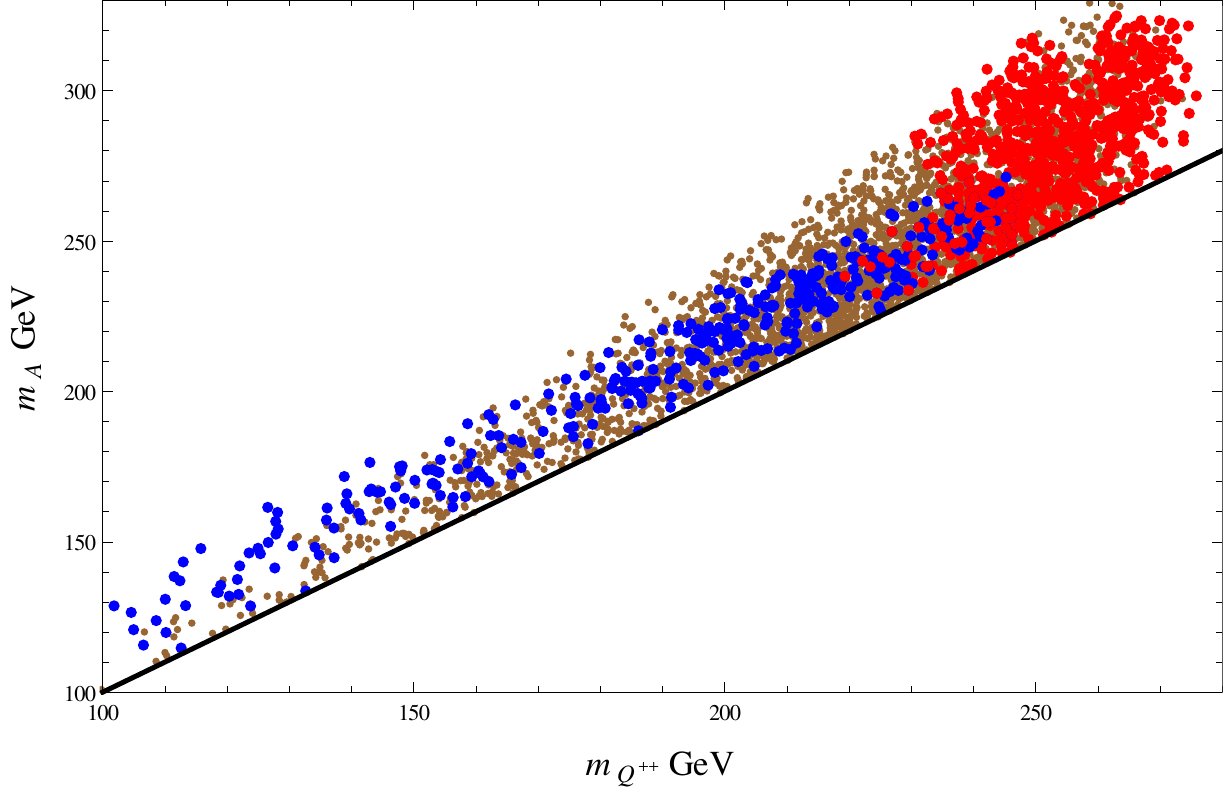}\hspace{0 cm} 
\includegraphics[width=8.5cm]{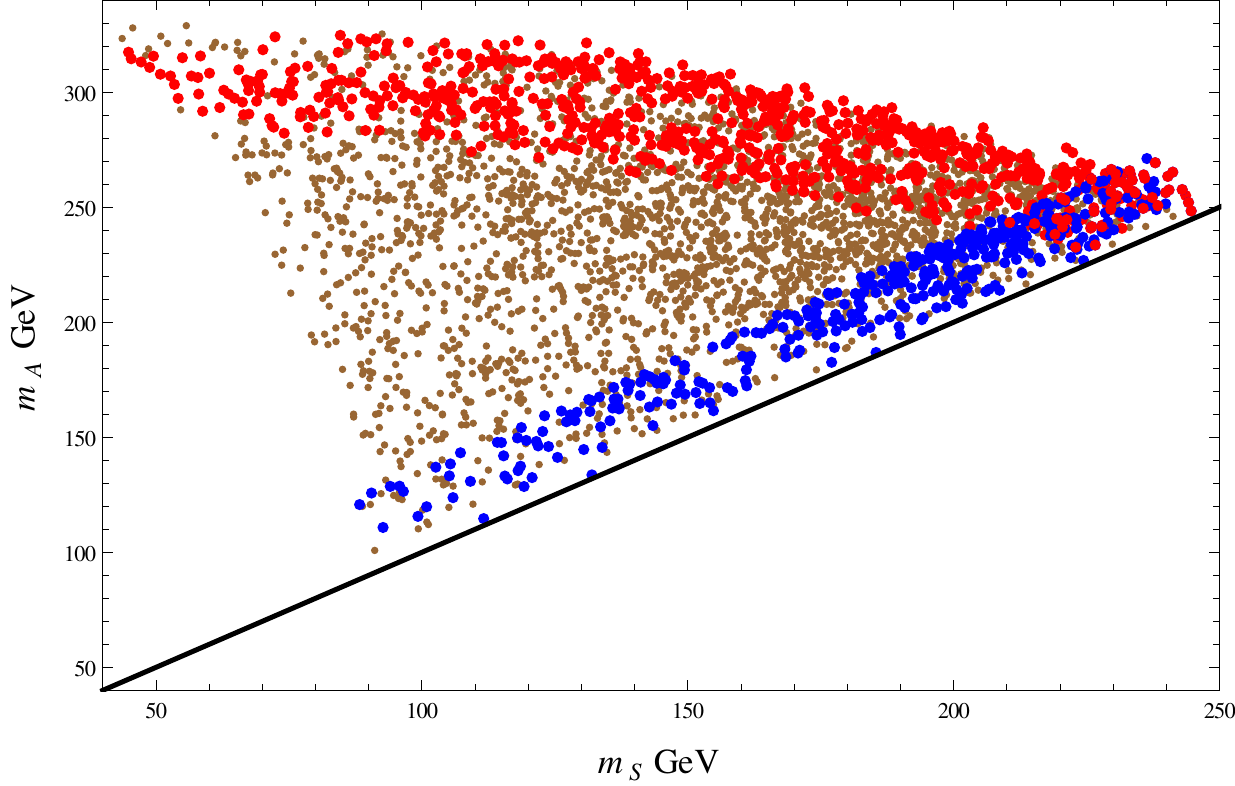}} 
\caption{Correlation between (left fig.) $m_{Q^{++}}$ and $m_{A}$ and 
(right fig.) $m_{S}$ and $m_{A}$. We have scanned the parameter space: $M\in
  (0,60)$ GeV, $\lambda_{2,3}\in (0,0.01)$, $\alpha\in (0, 2)$, $|\beta| \in (0,1.5)$ 
and $\gamma\in (0,1.5)$. Out of initial $10^5$ models, for $\beta>0$ (blue points in left and right fig.), $0.43\%$ points satisfy
stability conditions + precision data and mass hierarchy when
considering $S$ to be dark matter  
 where as for $\beta<0$ (brown points in both fig.), $3.47\%$ points satisfy the
same bounds. In addition, for $\beta>0$ and $\beta<0$, strong EWPhT
condition (red points in both figures) 
is satisfied by $0.04\%$ points with $m_{Q^{++}}\sim 200-250$ GeV and $m_{A}\sim 230-270$
and $0.8\%$ points with $m_{Q^{++}}\sim 230-275$ GeV and $m_{A}\sim 235-320$ respectively. The straight line in left fig. 
represents $m_{Q^{++}}=m_{A}$ and in right fig., $m_{S}=m_{A}$.}
\label{qrtptf}
\end{figure}

\paragraph{Expansion Parameter} 
One also has to keep in mind the validity of finite temperature perturbation expansion. 
In case of standard model, the first order phase transition is dominated by the gauge bosons 
but for the case of (previously considered) inert doublet, complex triplet or quartet, the 
phase transition is mainly driven by new scalar couplings to the Higgs. Therefore, we can 
safely neglect the gauge boson contribution. As an illustration, 
we can see by simplifying  Eq.(\ref{htpot}) and Eq.(\ref{htpotc}) that if the effective 
scalar coupling, $\alpha_{S}$ is responsible for the transition, in the region near the 
symmetry breaking minimum, $\phi/T \sim \alpha_{S}^{3/2}/\lambda_{1}$. The thermal mass of the 
corresponding particle is,
\begin{equation}
m^2_{S}(\phi,T)=M^2+\frac{1}{2}\alpha_{S}\phi^2+\Pi(T)
\label{tmm}
\end{equation}

Additional loop containing scalar will cost a factor $\sim\alpha_{S}T$ and loop expansion 
parameter can be obtained by dividing this factor with the leading mass of the theory 
which in this case is the mass of the new scalar; $\beta\sim \frac{\alpha_{S}T}{m_{S}}$. 
Now only in the limit, $M^2+\Pi(T)<<\alpha_{S}\phi^2$, 
we have, $\beta\sim\sqrt{\alpha_{S}}\frac{T}{\phi}$ or, near the region of 
minimum, $\beta\sim\frac{\lambda_{1}}{\alpha_{S}}\sim\frac{m^2_{h}}{m^2_{S}}$. 
Therefore, perturbation makes sense only for $\lambda_{1}<\alpha_{S}$ or $m_{h}<m_{S}$. 
On the other hand if $M^2+\Pi(T)$ term is significantly larger, it will reduce the 
order parameter and eventually transition will cease to be first order. In case of 
complex triplet, from Fig.(\ref{tript}), the 1st order EWPhT occurs 
for region: $m_{\Delta^{++}}\sim 170-340$ GeV. 
And for quartet, from Fig.(\ref{qrtptf}), first order 
EWPhT region is: $m_{Q^{++}}\sim 200-275$ GeV and $m_{A}\sim 230-320$ GeV. Therefore 
we can see that for both cases, the mass regions where first order transition 
occurs are larger than the Higgs mass and hence, within the validity of perturbation theory.

\subsection{Impact of Multiplets' Sizes on EWPhT}
\paragraph{Latent Heat Release\label{latentsec}}
The phase transition is characterized by the release of latent
heat. If there is a latent heat release, the transition is first order
in nature otherwise it is second or higher order (as in Ehrenfest's
classification).  The nature of cosmological phase transition is
addressed in \cite{Megevand:2003tg, Megevand:2007sv}. In this section,
we have addressed how the size of the representation affects the
latent heat release during the electroweak phase transition with
assumption that the transition is first order driven by large
Higgs-inert scalar coupling, i.e. $\alpha$. Consider a system gone
through first order phase transition at temperature, $T_c$. The high
temperature phase consists of radiation energy and false vacuum energy
and energy density is denoted as $\rho_{+}$. On the other hand,
although low temperature phase has equal free energy $F$ it will have
different energy density, $\rho_{-}$. The discontinuity
$\Delta\rho(T_c)=\rho_{+}(T_c)-\rho_{-}(T_c)$, gives the latent heat,
$L=\Delta\rho(T_c)=T_c\Delta s(T_c)$ where $\Delta s(T_c)=s_{+}-s_{-}$
is entropy density difference and it is liberated when the region of
high-T phase is converted into the low-T phase. Therefore, using $F=\rho-Ts$
with $s=-\frac{dF}{dT}$ and from expression for effective
potential (equivalent to free energy), Eq.(\ref{finited}) and as for 1st order 
phase transition, $F_{+}(0,T_c)=F_{-}(\phi_c,T_c)$, we have the latent heat for the
transition,

\begin{equation}
L=T\frac{dF_{-}}{dT}|_{(\phi_{c},T_{c})}-T\frac{dF_{+}}{dT}|_{(0,T_{c})}
\end{equation}

\begin{figure}[h!]
  \centerline{\includegraphics[width=9.5cm]{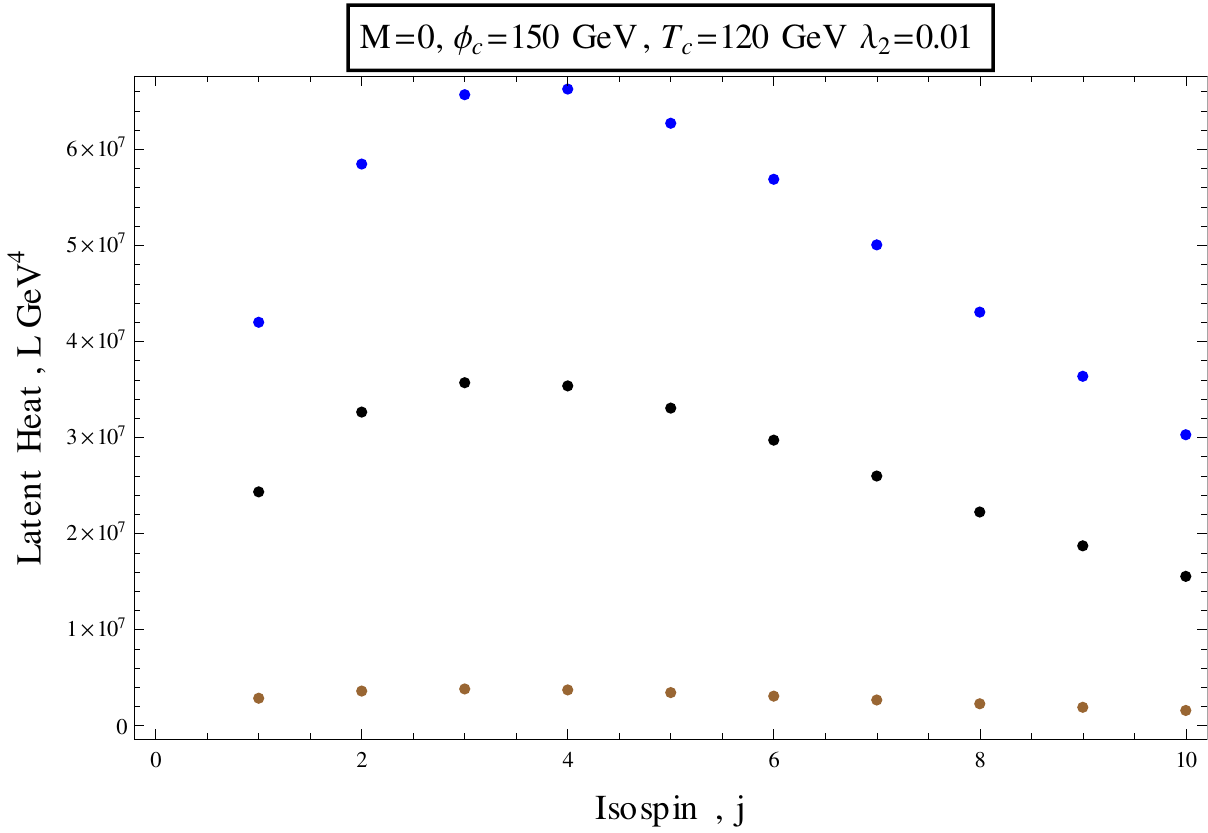}\hspace{0 cm}
    \includegraphics[width=9.5cm]{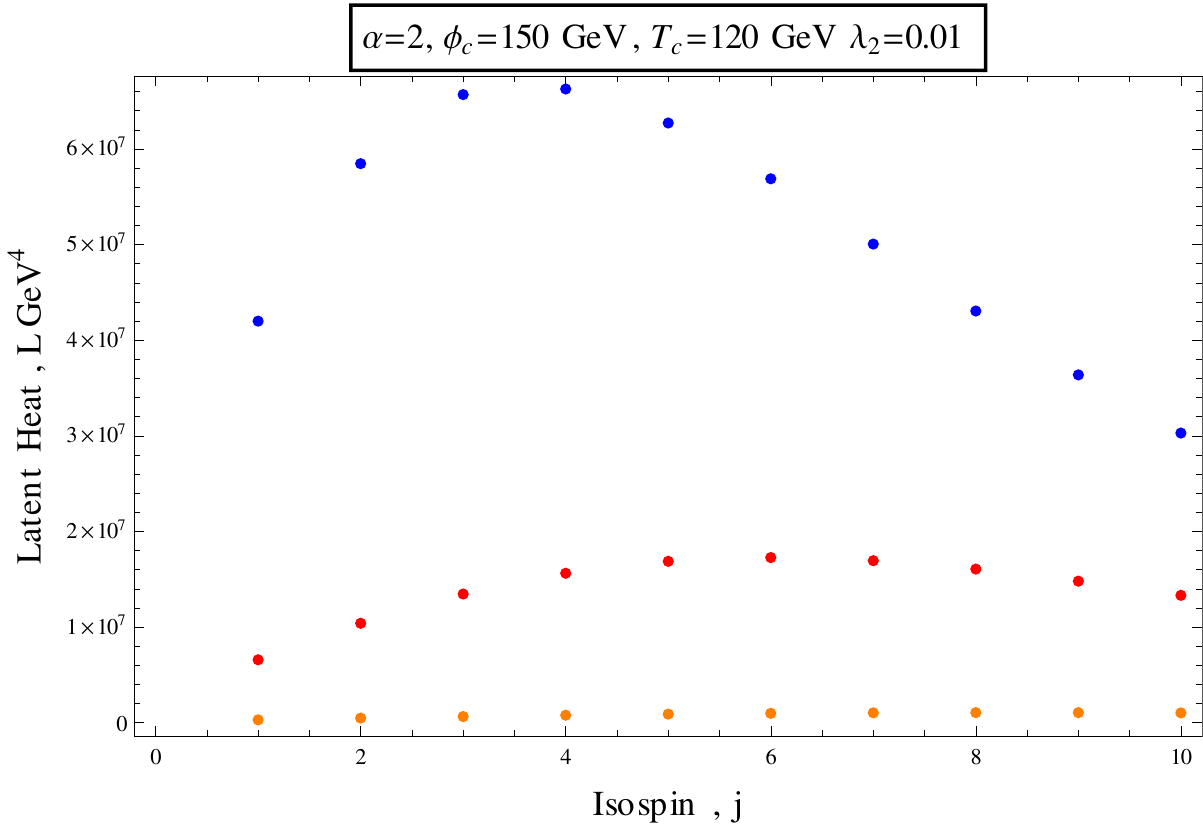}}
  \caption{In the left graph, latent heat decreases when isospin
    increases; blue, black and brown dots represents $\alpha=$2, $1$
    and $0.1$ respectively. On the right, we can see again the
    reduction of latent heat with isospin for invariant mass term,
    $M=0$ (blue), $500$ (red) and $1000$ (orange) GeV respectively.} 
  \label{latentf}
\end{figure}

For simplicity we can again take real degenerate
representation for probing the impact of dimension of large multiplet
on the latent heat release. As the amount of latent heat represents the strength of first order
transition, it is already clear from Fig.(\ref{latentf}) that larger
representation disfavors first order phase transition. In addition, in Sec.(\ref{rgeq}) 
we have seen that arbitrarily large scalar multiplet makes gauge and scalar couplings 
non-perturbative in TeV or even at smaller scale. Therefore, larger scalar multiplet 
cannot simultaneously strengthen the electroweak phase transition and stay consistent 
with perturbativity and unitarity of the theory. Similar conclusion can be drawn for 
complex even integer and half integer
multiplets.

\section{The Quartet/Doublet versus EW, EWPhT and CDM
  Constraints. \label{ldm}}  
In this section, we have tried to identify region of parameter space
for higher representations where one can have light DM candidate
consistent with other phenomenological constraints along with
strong EWPhT. As already pointed out in  Sec.(\ref{tripletsec}), DM of
complex even integer multiplet ($Y\neq 0$) is excluded by the bound 
from direct detection. On the other hand,
the $\gamma$ term of Eq.(\ref{potq}) can split the neutral component of
half integer representation and one can easily obtain a light DM
component. As quartet multiplet $(j=3/2)$ is the immediate generalization of inert 
doublet, we focused on identifying DM properties in parallel with it's impact on Strong EWPhT.

 \subsection{Model Parameters Scan and Constraints}
The masses of quartet's component fields are determined by four free
parameters $\{M_{Q},\alpha,\beta,\gamma\}$, Eq.(\ref{qrmass1}). But (co)annihilation cross
  sections which control the 
relic density of the dark matter depend on the mass splittings among
dark matter and the others components in the multiplet. Therefore we
used alternative free parameter 
set $\{m_{S},\lambda_{S},m_{Q^{++}},m_{A}\}$ where
$m_{S}$ is the DM mass and 
$\lambda_{S}=\alpha+\frac{1}{4}\beta-2\gamma$ is the coupling between
between Higgs and dark matter component. One
can express $\{M_{Q},\alpha,\beta,\gamma,
m_{Q^{+}_{1}},m_{Q^{+}_{2}}\}$ in terms of these four parameters using 
mass relations Eq.(\ref{qrmass1}). Moreover, in case of quartet, for $S$ to be the lightest component of
the multiplet, one needs to impose two conditions, $\gamma > 0$ and $\gamma \geq
\frac{|\beta|}{2}$ which in turn sets the mass spectrum to be 
$m_{S}<m_{Q^{+}_{1}}<m_{Q^{++}}<m_{Q^{+}_{2}}<m_{A}$.
\\

{\bf Collider Constraints:}\\
Direct collider searches at LEP II has put a strong bound on single
charged particle which is $m_{Q^{+}_1}>70-90$ GeV \cite{lep}. But the
doubly charged scalar of the quartet which only has the
cascade decay channel is not strongly constrained by collider
searches. One constraint can come from $W$ and $Z$ boson width. In
our case, setting $S$ as DM imposes in the mass spectrum: $m_{Q^{++}}
\geq m_{Q^{+}_{1}}$ so the constraint on single charged scalar is also
translated into a bound on the doubly charged scalar for such mass
spectrum. Moreover, the deviations of $W$ and $Z$ width from their SM
values can take place through decay channels: $W^{\pm}\rightarrow
SQ^{\pm}_{1}/AQ^{\pm}_{1}/Q^{\pm\pm}Q^{\mp}_{1}$ and $Z\rightarrow
Q^{+}_1Q^{-}_1/SA/Q^{++}Q^{--}$. Therefore to avoid such deviation the
following mass constraints are also imposed: $m_S+m_{Q^{+}_{1}}>m_W$,
$m_A+m_{Q^{+}_{1}}>m_W$, $m_{Q^{++}}+m_{Q^{+}_{1}}>m_W$,
$m_{Q^{+}_{1}}>m_Z/2$, $m_{Q^{++}}>m_Z/2$ and $m_S+m_A>m_Z$. Apart
from collider constraints, one also impose constraints coming from
electroweak precision observables. In our scan, we used the allowed
range of $T$ parameter, Eq.(\ref{tp}) and $S$ parameter,
Eq.(\ref{sp}).
\\

{\bf DM Relic Density Constraint:}\\ 
The dark matter density of the universe measured by Planck
collaboration is $\Omega_{DM}h^2=0.1196\pm 0.0031$ ($68\%$ CL)
\cite{Ade:2013zuv}. To determine the relic density of the multiplets, we used FeynRules
\cite{Christensen:2008py} to generate the model files for MicrOMEGAs
\cite{Belanger:2010gh}. For inert multiplets, mass
splitting between components are set by both $\beta$ and $\gamma$
couplings. In case of doublet, one can set $\beta$ and $\gamma$ to
produce large spitting between $S$ component and single charged
component $C^{+}$ or between $S$ and $A$ in such way that such
splitting is compatible with electroweak precision observables. Such
large splitting can lead to suppression of co-annihilation channels
$SA, SC^{+}\rightarrow$ SM particles. But such simple tuning of the
couplings like in the doublet is not possible for quartet because of
the mass relation Eq.(\ref{qrmass1}). Therefore there is a possibility for
co-annihilation channel to open up when $m_{Q^{+}_1}/m_{S} \leq 1.5$
(\cite{Griest:1990kh}).

Apart from co-annihilation, other dominant channels which will control
the relic density of DM at the low mass region are $SS\rightarrow 
h^{*}\rightarrow b\bar{b}$ and $SS\rightarrow WW$.
\\

{\bf Direct DM Detection and Invisible Higgs Decay Constraints:}\\
There is also a strict limit on the spin independent DM-nucleon cross
section coming from direct detection experiment. The spin independent
cross section is given by,  
\begin{equation}
  \sigma_{SI}=\frac{\lambda_{S}^2 f^2}{4\pi} \frac{\mu^2 m^2_{n}}{m^4_h m^2_S}
\end{equation}
Here, $\mu=m_n m_s/(m_n+m_s)$ is the DM-nucleon reduced mass. $f$
parameterizes the nuclear matrix element, $\sum_{u,d,s,c,b,t}\langle
n|m_q \bar{q}q|n\rangle\equiv f m_n \bar{n}{n}$ and from recent
lattice results \cite{Giedt:2009mr}, $f=0.347131$. Now XENON100 with
225 days live data \cite{Aprile:2012nq} sets bound on cross section to
be $\sigma_{SI} \lesssim 2 \times 10^{-45}\text{cm}^2$. In addition,
the future XENON1T will reach the sensitivity of
$\sigma_{SI} \lesssim 2 \times 10^{-47}\text{cm}^2$ \cite{Aprile:2012zx}.

Also if the mass of the dark matter is smaller than half of the
Higgs mass, it will contribute
to the invisible decay of the Higgs through $h\rightarrow SS$ with branching ratio
$Br_{\text{inv}}=\Gamma_{\text{inv}}/(\Gamma_{\text{SM}}+\Gamma_{\text{inv}})$,
where $\Gamma_{\text{SM}}=6.1$ MeV and,
\begin{equation}
 \Gamma_{inv}=\frac{\lambda_{S}^2 v^2}{64\pi m_{h}}\sqrt{1-\frac{4 m_{S}^2}{m_{h}^2}}
\end{equation}

\begin{figure}[h!]
  \centerline{\includegraphics[width=8cm]{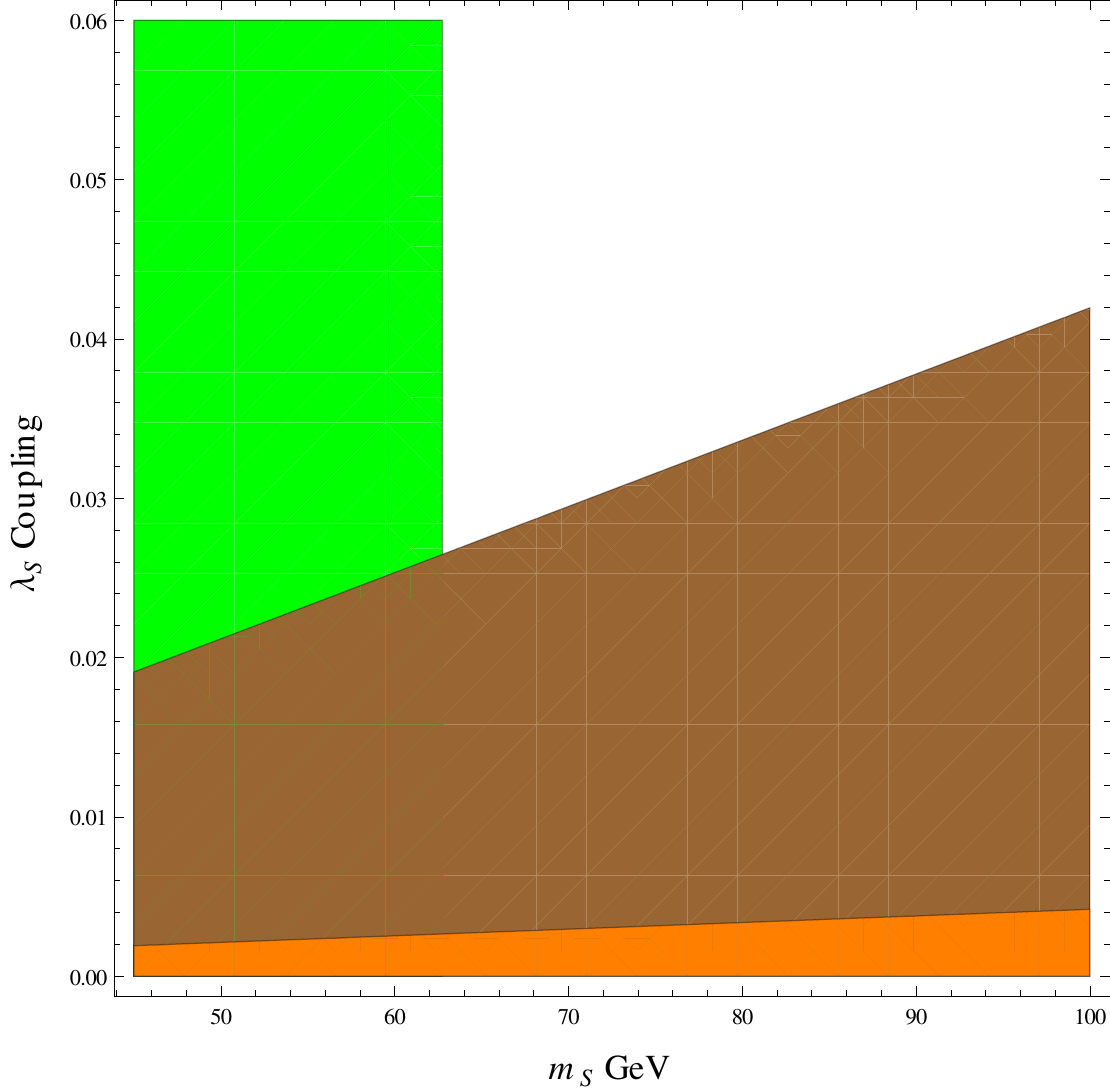}} 
  \caption{The region in $m_{S}-\lambda_{S}$ plane allowed by invisible
branching ratio limit $Br_{inv}\leq 0.65$ (green region) and XENON100 limit (brown region) and 
XENON1T limit (orange region) respectively for DM mass range $45 \leq m_{S}\leq 100$ GeV. The future 
XENON1T will significantly reduce the allowed region down to $0.002\lesssim \lambda_{S} \lesssim 0.004$
for $m_{S}=45-100$ GeV.}  
\label{direct}
\end{figure}

Consequently, current limit on the
branching ratio for Higgs invisible decay \cite{invhiggs} will
constrain the $\lambda_{S}$ coupling. However, note that as can be
seen in Fig.(\ref{direct}), the limit imposed on $\lambda_{S}$  by
XENON100 is more stringent than the limit coming from invisible Higgs
decay for mass range $45 \leq m_S \leq 63$  GeV.

\subsection{Allowed Parameter Regions}
To determine allowed parameter region for quartet and compare it with
inert doublet case, we used dark matter relic density constraint at
$5\sigma$ to take into account all the numerical
uncertainties. For numerical scanning, we considered the following range: $m_{S}\in
(45,100)$ GeV, $\lambda_{S} \in (0.001,0.02)$, $m_{Q^{++}} \in
(100,350)$ GeV and $m_{A} \in (100,350)$ GeV for both doublet and quartet case.

At low mass region $(m_{S}\leq m_{W})$ the dominant annihilation channel is
$SS\rightarrow h^{*}\rightarrow b\bar{b}$. When the mass becomes
larger than $m_{W}$ and $m_{Z}$, $SS\rightarrow W^{+}W^{-}$ and
$SS\rightarrow ZZ$ open up and dominate the annihilation
rate. Therefore the relic density becomes much lower than the observed
relic density. Subsequent increase of $m_S$ will open up
$SS\rightarrow hh$ and $SS\rightarrow t\bar{t}$ which will dominate
along with $WW$ and $ZZ$ annihilation channels and eventually the
relic density will be much smaller than observed value. Inert doublet
has already shown such behavior \cite{Dolle:2009fn, Melfo:2011ie}. 
	
\begin{figure}[th]
  \centerline{\includegraphics[width=8.5cm]{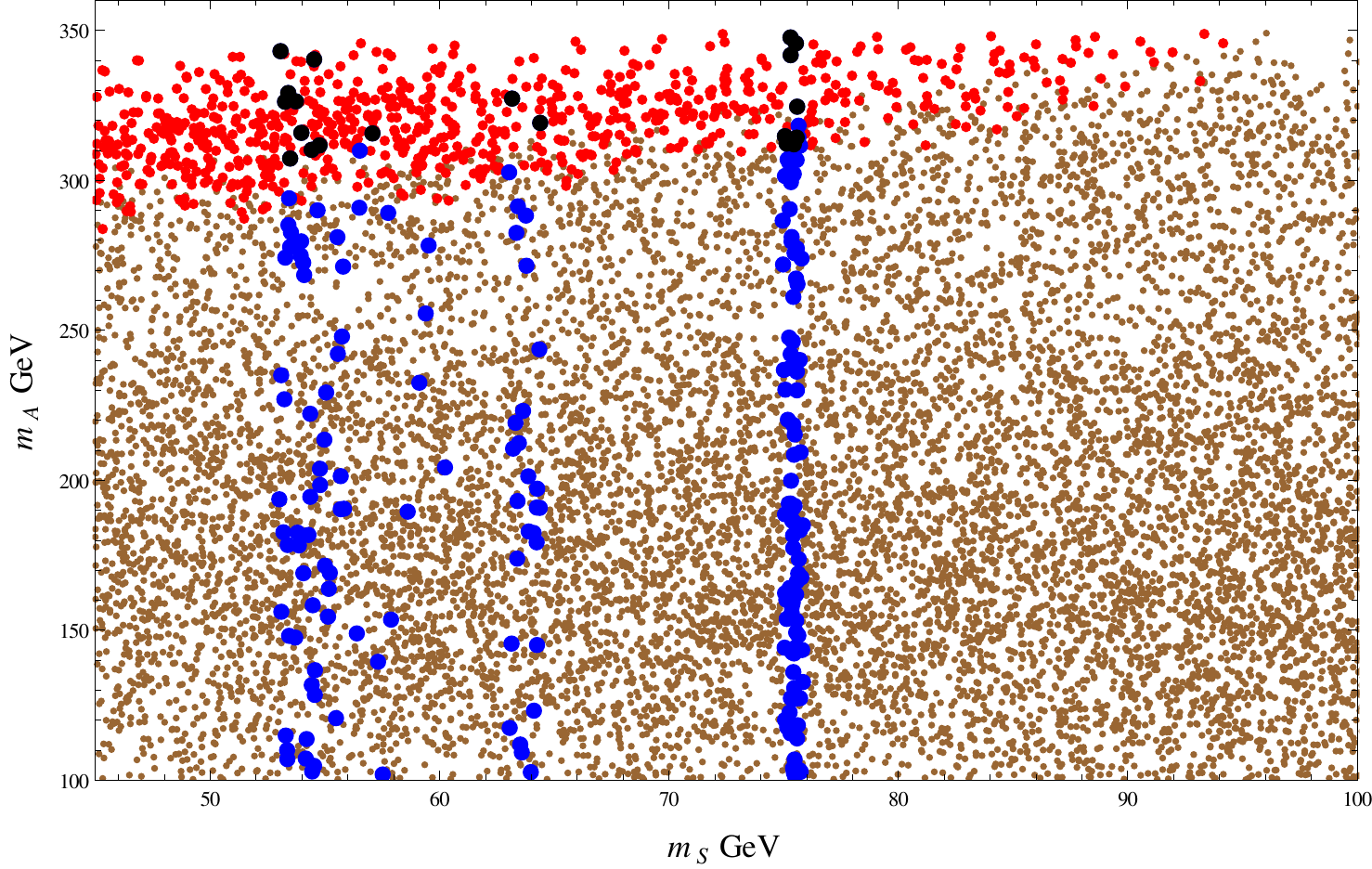}\hspace{-0.1cm} 
    \includegraphics[width=8.5cm]{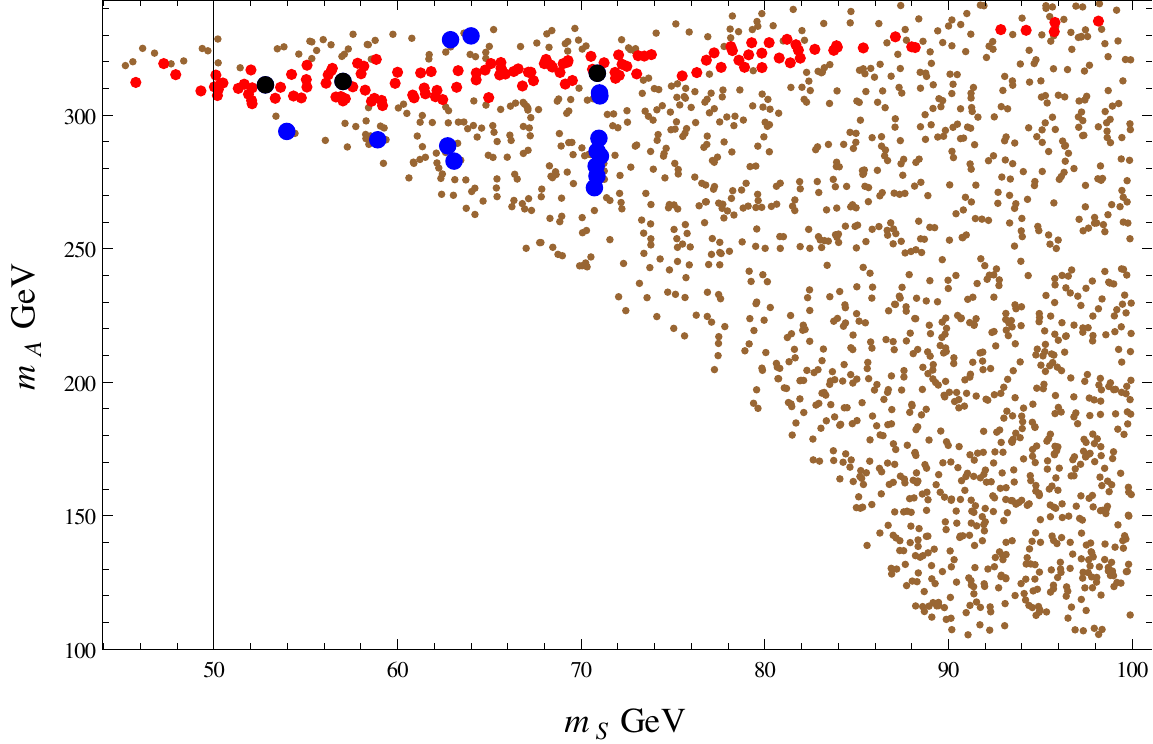}}
  \caption{Scatter plot representing the correlation between $m_S$ and
    $m_A$ for doublet(left fig.) and quartet (right fig.). Brown
    points represent models satisfying stability conditions + precision
    bounds+collider constraints. Red points correspond to models also
    satisfying the EWPhT condition $\phi_c/T_c
    \geq 1$ and blue points correspond to models consistent only with
    observed relic density at $5\sigma$ and DM direct detection and
    invisible Higgs decay bounds. Black points are models 
    consistent with stability conditions + EWPD + collider constraints
    + the EWPhT condition $\phi_c/T_c\geq 1$ + observed $\Omega_{DM}h^2$
    + DM direct detection and invisible Higgs decay limits.}
  \label{qrel} 
\end{figure}

From scatter plot, Fig.(\ref{qrel}) we can see that, unlike the 
doublet case, for quartet, the parameter space allowing both light DM 
consistent with observed relic density and strong first order phase 
transition is hard to achieve. Out of an initial of $10^5$ models, for
doublet, $20\%$ are consistent with stability conditions + precision
data+ collider constraints, $0.77\%$ satisfy,
in addition, the EWPhT condition $\phi_c/T_c\geq 1$ and $0.234\%$
agree with the 
observed $\Omega_{DM}h^2$ at $5\sigma$, DM direct detection bounds and
invisible Higgs decay limits. Only $0.02\%$ of
the initial models survive all the constraints. In contrast, for
quartet, from an initial of $10^5$ models only $2\%$ satisfy stability
conditions + EWPD +collider constraints, $0.13\%$ points satisfy additional strong EWPhT condition and only 
$0.03\%$ models, observed $\Omega_{DM}h^2$ at $5\sigma$ and direct detection bounds. Lastly,
only $0.003\%$ models satisfied all the conditions. Therefore, in case of providing
strong EWPhT with a light dark matter, quartet is disfavored with respect to doublet.

\section{Conclusions and outlook} \label{con}
We have considered the electroweak phase transition and dark matter
phenomenology with various inert scalar representations used for
extending the SM's Higgs sector. The details of the phenomenological
studies are done by making random scans of parameters within the
triplet and quartet context. The results from our analyses in
comparing the allowed parameter regions of the above mentioned models
with the inert doublet model case are summarized as follows

\begin{itemize}
\item As the size of an inert
multiplet which can be added to the SM is not arbitrary but is rather 
controlled by the perturbativity of SU(2) gauge coupling at TeV scale ($\sim 10$ TeV),
which sets an upper bound on the size of the multiplet to be $j\leq 5$, 
that motivated us to study the EWPhT (and subsequently, the DM characteristics) for the triplet and
quartet models as representatives for allowed larger multiplets. 
We explicitly showed that it is possible to have strong EWPhT within the complex
inert triplet and the inert quartet models. In case of
complex triplet, from Fig.(\ref{tript}), the 1st order EWPhT occurs 
for region: $m_{\Delta^{++}}\sim 170-340$ GeV. 
And for quartet, from Fig.(\ref{qrtptf}), first order 
EWPhT region is: $m_{Q^{++}}\sim 200-275$ GeV and $m_{A}\sim 230-320$ GeV
as compared to the mass of inert doublet's singly charged component, $m_{C}$
($\sim$ pseudo-scalar component's mass, $m_{A}$)
in $270-350$ GeV \cite{Chowdhury:2011ga}.

\item Using the expression for the
  latent heat measure, we have made a generic study of
  the impact of higher (than the doublet) scalar multiplets on the strength of electroweak phase
  transition. The compete between the number of scalar quasiparticles
  coupled to EW plasma with large couplings and the screening of those particles
  resulting from scalar self quartic and gauge boson interaction, which will
  decouple them from the plasma, determines the 
  strength of the transition. The rise and fall of the latent heat with large 
  multiplet, as shown in Fig.(\ref{latentf}), qualitatively shows  
  such variation of the EWPhT strength.

\item Next we require the simultaneous explanation of EWPhT and the 
  cold DM content of the universe within the quartet frame. The
  triplet with $Y=2$, unlike the half integer representations, is already
  excluded by the direct detection limit thus can't play any viable role of DM
  triggering strong EWPhT. Therefore for quartet, we identified the region of
  parameter space, out of the randomly scanned parameter points, which
  allows both strong EWPhT and a light DM candidate by imposing stability
  conditions, electroweak precision bounds and corresponding experimental limits.
  The parameter regions that survive all constraints imposed
  are compared with similar results from within the inert doublet
  frame. Requiring that all the CDM content of the universe is
  explained by the scalar multiplet then, from the scatter plot
  Fig.(\ref{qrel}), it can be seen that the quartet has only a very
  small allowed parameter space in contrast 
  to inert doublet case. Moreover, the allowed parameter space
  for having DM with observed relic density and strong EWPhT in 
  both inert doublet and quartet cases will be significantly constrained 
  by future XENON1T experiment.
\end{itemize}

The above results combined together point out that the higher inert
representations are rather disfavored compared to the inert
doublet. The conclusion is qualitative and only valid within the set
of experimental and phenomenological constrains considered. There are
however room for doing more in directions we did not consider here:
\begin{itemize} 
\item The computational and phenomenological machineries are well
  within reach for making detailed quantitative analyses within 
  Bayesian statistical framework. Using strong EWPhT condition and recent results from 
  relic density measurement, direct and indirect detection limits and collider constraints from LHC,
  a comparison between the allowed parameter spaces of inert doublet and quartet can be carried out 
  in this Bayesian framework, in a similar manner to the work done for comparing 
  different supersymmetric models \cite{Feroz:2008wr,AbdusSalam:2009tr} and 
  for comparing a single versus multi-particle CDM universe hypotheses
  \cite{AbdusSalam:2010qp}. 

\item Higher (than doublet) scalar multiplets have relatively more
  charged components that couple to the Higgs. As such, they will
  alter the decay rate of Higgs going to two photons relative to the
  SM value \cite{Ellis:1975ap, Shifman:1979eb, Carena:2012xa, Fan:2013qn,  Picek1, Brdar:2013iea}. 
  Persistence of the apparent (given the large
  uncertainties) excess in the $h\rightarrow \gamma\gamma$ data
  will severely constrain the triplet and quartet parameter regions
  with EWPhT driven by large positive couplings \cite{ Chung:2012vg, Huang:2012wn}. 
  This and similar studies will be interesting for establishing the 
  inert multiplets' status and/or prospects within collider phenomenology framework.

\end{itemize}

\noindent{\bf Acknowledgements:}\\
\\
T.A.C. is deeply indebted to Goran Senjanovi\'c for the suggestion to
investigate larger scalar multiplets and the guidance in every step of
the work. T.A.C. is also grateful to Yue Zhang and Miha Nemev\v sek
for many helpful discussions and Amine Ahriche for careful reading of
the manuscript. T.A.C. would also like to thank Andrea de Simone,
Marco Serone, Gabrijela Zaharijas, Abdesslam Arhrib, Rachik Soualah,
Muhammad Muteeb Nouman and Arshad Momen for helpful comments and
discussions.

\section{Appendix}

\subsection{Quartet $SU(2)$ Representation Generators \label{apa}}

The generators $R(T^a)$ in representation $R$ of $SU(2)$ are taken in
such a way that they satisfy the following relation:
$Tr[R(T^a)R(T^b)]=T(R) \delta^{ab}.$ Here $T(R)$ is the dynkin index
for the corresponding representation. It is obtained from $T(R)
D(Ad)=C(R) D(R)$ where dimension of Adjoint reps. is $D(Ad)=3$ for
$SU(2)$ and Casimir invariant is $\sum_a R(T^a)R(T^a)=j(j+1)$ and
dimension of the reps. $R$ is $D(R)=2j+1$. For $SU(2)$
reps. $T(\frac{1}{2})=\frac{1}{2}$ and $T(\frac{3}{2})=5$. 

The explicit form of the generators for quartet reps are the
following: 

\begin{equation}
  T^1=\begin{pmatrix}
    0&\frac{\sqrt{3}}{2}&0&0\\
    \frac{\sqrt{3}}{2}&0&1&0\\
    0&1&0&\frac{\sqrt{3}}{2}\\
    0&0&\frac{\sqrt{3}}{2}&0\\
  \end{pmatrix}\,, \quad 
  T^2=\begin{pmatrix}
    0&-\frac{\sqrt{3}i}{2}&0&0\\
    \frac{\sqrt{3}i}{2}&0&-i&0\\
    0&i&0&-\frac{\sqrt{3}i}{2}\\
    0&0&\frac{\sqrt{3}i}{2}&0\\
  \end{pmatrix}\,, \quad 
  T^3=\text{diag}(\frac{3}{2},\frac{1}{2},-\frac{1}{2},-\frac{3}{2}). 
\end{equation}
The raising and lowering operators are defined as $T^{\pm}=T^1\pm
iT^2$. The antisymmetric matrices in doublet reps is $\epsilon$ and
the similar antisymmetric matrix for quartet representation, $C$ is
constructed using the relation $CT^aC^{-1}=-T^{aT}$ because the
explicit form of it depends on the matrix representation of the
corresponding generators:
\begin{equation}
\epsilon=\begin{pmatrix}
    0&1\\
    -1&0\\
  \end{pmatrix}\,, \quad 
  C=\begin{pmatrix}
    0&0&0&1\\
    0&0&-1&0\\
    0&1&0&0\\
    -1&0&0&0
  \end{pmatrix}
\end{equation}

\subsection{High Temperature Expansion of Thermal
  Potential \label{ahigheff}}
  
The finite temperature effective potential is
\begin{eqnarray}
  V_{T}&=&\sum_{B(F)}(\pm) g_{B(F)}\frac{T^4}{2 \pi^2}\int_0^\infty
  dxx^2\ln(1\mp e^{-\sqrt{x^2+m_{B(F)}^2(\phi_i,T)/T^2}})\nonumber\\
  &=& \sum_{B(F)}(\pm) g_{B(F)}\frac{T^4}{2 \pi^2}J_{B(F)}(m_{B(F)}^2(\phi_i,T)/T^2)
\end{eqnarray}

where, $J_{B}(m^2_{B}/T^2)$ and $J_{F}(m^2_{F}/T^2)$ are denoting bosonic and fermionic integral respectively.
 
In the High temperature limit, $m(\phi)/T<1$, finite temperature bosonic integral is
\begin{eqnarray}
  \label{htb}
 J_B(\frac{m^2}{T^2}) &=& \int_0^\infty dxx^2\ln(1-
  e^{-\sqrt{x^2+m^2/T^2}}) =
  -\frac{\pi^4}{45}+\frac{\pi^2}{12}\frac{m^2}{T^2} -
  \frac{\pi}{6}(\frac{m^2}{T^2})^{\frac{3}{2}} -
  \frac{1}{32}\frac{m^4}{T^4}\ln\frac{m^2}{A_{b}T^2} \\  \nonumber   
  &-&2\pi^{\frac{7}{2}}\sum_{l=1}^{\infty}(-1)^l\frac{\zeta(2l+1)}{(l+1)!}\Gamma(l+\frac{1}{2})(\frac{m^2}{4\pi^2T^2})^{l+2}
  \quad \textrm{where } \, A_b=16 \pi^2 e^{(3/2-2 \gamma_E)} (\ln A_b=5.4076),
\end{eqnarray}
$\gamma_E$ is Euler's constant and $\zeta$ is the Riemann
$\zeta$-function. The fermionic integral is 
\begin{eqnarray}
  \label{htf}
 & & J_F(\frac{m^2}{T^2}) = \int_0^\infty dxx^2\ln(1+ 
  e^{-\sqrt{x^2+m^2/T^2}}) =
  -\frac{7\pi^4}{360}+\frac{\pi^2}{24}\frac{m^2}{T^2} -
  \frac{1}{32}\frac{m^4}{T^4}\ln\frac{m^2}{A_{f}T^2}\\ \nonumber   
  &-&\frac{\pi^{\frac{7}{2}}}{4}\sum_{l=1}^{\infty}(-1)^l\frac{\zeta(2l+1)}{(l+1)!}(1-2^{-2l-1})\Gamma(l+\frac{1}{2})(\frac{m^2}{4\pi^2T^2})^{l+2}
  \quad \textrm{where } \, A_f= \pi^2 e^{(3/2-2 \gamma_E)} (\ln A_b=2.6351).
\end{eqnarray}

\subsection{Renormalization Group Equations\label{rgeqn}}

The doublet RG equations \cite{Barbieri:2006dq} in our parameterization are in the following. Here $g_{2}$, $g_{Y}$ and $g_{3}$ are $SU(2)_{L}$, $U(1)_{Y}$ and $SU(3)_{c}$ couplings respectively.

\begin{eqnarray}
16\pi^2\beta_{\lambda_{1}} &=& 24 \lambda_1^2+2 \alpha^2+\frac{1}{8}\beta^2+\gamma^2-9\lambda_{1}g_{2}^2-3\lambda_{1}g_{Y}^2+\frac{9}{4}g_{2}^4+\frac{3}{4}g_{Y}^4+\frac{3}{2}g_{2}^2 g_{Y}^2+12\lambda_{1} y_{t}^2-6y_{t}^4\nonumber\\
16\pi^2\beta_{\lambda_{2}} &=& 24 \lambda_2^2+2 \alpha^2+\frac{1}{8}\beta^2+\gamma^2-9\lambda_{2}g_{2}^2-3\lambda_{2}g_{Y}^2+\frac{9}{4}g_{2}^4+\frac{3}{4}g_{Y}^4+\frac{3}{2}g_{2}^2 g_{Y}^2\nonumber\\
16\pi^2\beta_{\alpha} &=& 4\alpha^2+12 \lambda_1 \alpha+\lambda_2 \alpha+\frac{3}{4}\beta^2+6 \gamma^2-9\alpha g_{2}^2-\alpha g_{Y}^2+\frac{9}{4}g_{2}^4+\frac{3}{4}g_{Y}^4-\frac{3}{2}g_{2}^2
g_{Y}^2+12\alpha y_{t}^2\nonumber\\
16\pi^2\beta_{\beta} &=& 4\lambda_1 \beta+4 \lambda_2 \beta+8 \alpha \beta +16 \gamma^2-9\beta g_{2}^2-\beta g_{Y}^2+12\beta  y_{t}^2\nonumber\\
16\pi^2\beta_{\gamma} &=& 4 \lambda_1 \gamma+4 \lambda_2 \gamma+8 \alpha \gamma+4 \beta \gamma-9\gamma g_{2}^2-\gamma g_{Y}^2+12\gamma  y_{t}^2\nonumber\\
16\pi^2\beta_{g_{2}}&=&-3g_{2}^3,\,\,\, 16\pi^2\beta_{g_{Y}}=\frac{22}{3}g_{Y}^3,\,\,\, 16\pi^2\beta_{g_{3}}=-7g_{3}^3\nonumber\\
16\pi^2\beta_{y_{t}}&=&y_{t}(\frac{9}{2}y_{t}^2-\frac{17}{12}g_{Y}^2-\frac{9}{4}g_{2}^2-8 g_{3}^2)
\end{eqnarray}

And the Triplet RG equations \cite{Schmidt:2007nq} relevant for our analysis are,

\begin{eqnarray}
\label{rgt}
16\pi^2\beta_{\lambda_{1}} &=& 24 \lambda_1^2+6 \alpha^2+\beta^2-9\lambda_{1}g_{2}^2-3\lambda_{1}g_{Y}^2+\frac{9}{2}g_{2}^4+\frac{3}{2}g_{Y}^4+3 g_{2}^2 g_{Y}^2+12\lambda_{1} y_{t}^2-6y_{t}^4\nonumber\\
16\pi^2\beta_{\lambda_{2}} &=& 28\lambda_2^2+48\lambda_{2}\lambda_{3}+24\lambda_{3}^2+2 \alpha^2+\beta^2-24\lambda_{2}g_{2}^2-12\lambda_{2}g_{Y}^2+6g_{2}^4+6 g_{Y}^4+24 g_{2}^2 g_{Y}^2\nonumber\\
16\pi^2\beta_{\lambda_{3}} &=& 36\lambda_3^2+24\lambda_{2}\lambda_{3}-\frac{1}{2}\beta^2-24\lambda_{3}g_{2}^2-12\lambda_{3}g_{Y}^2+3g_{2}^4-12 g_{2}^2 g_{Y}^2\nonumber\\
16\pi^2\beta_{\alpha} &=& 3\alpha^2+6 \lambda_1 \alpha+16\lambda_2 \alpha+24\lambda_{3}\alpha-\frac{33}{2}\alpha g_{2}^2-\frac{15}{2}\alpha g_{Y}^2+6 g_{2}^4+3 g_{Y}^4+12 y_{t}^2\alpha\nonumber\\
16\pi^2\beta_{\beta} &=& 2\lambda_1 \beta+4 \lambda_2 \beta-4\lambda_{3}\beta+16 \alpha \beta-\frac{33}{2}\beta g_{2}^2-\frac{15}{2}\beta g_{Y}^2+12 g_{2}^2 g_{Y}^2-6\beta  y_{t}^2\nonumber\\
16\pi^2\beta_{g_{2}}&=&-\frac{5}{2}g_{2}^3,\,\,\, 16\pi^2\beta_{g_{Y}}=\frac{47}{6}\sqrt{\frac{5}{3}}g_{Y}^3,\,\,\, 16\pi^2\beta_{g_{3}}=-7g_{3}^3\nonumber\\
16\pi^2\beta_{y_{t}}&=&y_{t}(\frac{9}{2}y_{t}^2-\frac{17}{12}g_{Y}^2-\frac{9}{4}g_{2}^2-8 g_{3}^2)
\end{eqnarray}

\end{document}